\documentclass{desyproc}

\def\cascade{{\sc Cascade}}
\def\pythia{{\sc Pythia}}
\def\herwig{{\sc Herwig}}
\def\powheg{{\sc Powheg}}
\def\mcatnlo{{\sc MC@NLO}}

\begin{document}


\title{ \hspace*{2.2 cm }  Soft Gluons and  Jets  at the LHC\\ 
}



%
%
%
%
\author{\hspace*{6.3 cm }   {\slshape F.~Hautmann}\\[1ex]
\hspace*{2.5 cm } Theoretical Physics, 
 University of Oxford, 
  Oxford OX1 3NP \\
\hspace*{2.2 cm }  Physics  and  Astronomy, 
 University of Sussex,   Brighton   BN1 9QH 
   }
%
%
%
%
%



\contribID{ZZ}
\confID{UU}
\desyproc{DESY-PROC-2012-YY}
\maketitle


\begin{abstract}
We   address    aspects of jet physics at the Large Hadron Collider  focusing on   
features of recent   jet measurements  which challenge the theory.  
We  discuss  examples illustrating  the role of 
QCD parton  showers, nonperturbative corrections, 
soft multi-gluon  emission.

\vskip 0.3cm 
\hspace*{0.1cm} Based on talks given  at  the conferences   
{\it  Physics at the LHC 2011} (Perugia,  June 2011),   \\    
\hspace*{0.1cm}  
{\it  QCD at Cosmic Energies - V} (Paris, June 2012),  
 {\it  ISMD 2012} (Kielce, September 2012)
\end{abstract}



\vskip 0.8 cm

\section{Introduction}

The first three years of running of the LHC have  probed 
 jet physics in new ways, investigating 
previously unexplored kinematic regions.  
While next-to-leading-order (NLO) QCD calculations,  
supplemented with 
nonperturbative corrections and  parton showers,
are able to describe  well 
inclusive jet   spectra over a wide range of transverse momenta  
 extending   from 20 GeV to 2 TeV,  
several   features  of  LHC jet data  challenge the theory. 
 This applies in particular to    
 the behavior of  cross sections  with increasing rapidity;      to    correlations of multiple 
jets   in rapidity,  azimuthal angle,  transverse energy; 
 to    non-inclusive  observables  probing the  structure 
  of   high multiplicity final  states.   

This  article    focuses on  aspects of jet production      which,  
 despite the presence of a  high transverse momentum scale, 
are  sensitive to soft gluon processes and  QCD  infrared physics.  
We  start  with inclusive  cross sections in 
Sec.~\ref{sec:jet}  and  discuss  the role 
 of    parton showering  and   nonperturbative effects 
 in the   context of matched NLO-shower event 
generators. In Sec.~\ref{sec:forw}  we consider  forward jets, and      
examples  of   multi-jet  correlations.    
   Sec.~\ref{sec:heavy}   examines    $b$-flavor jets. 
   Sec.~\ref{sec:under}   takes a further look at   jet correlations from the 
viewpoint of   multiple parton interactions, emphasizing the 
 role    of   energy flow variables.  
 Sec.~\ref{sec:mini} addresses     motivation  and prospects for 
    extending   jet measurements 
     to lower transverse momenta than is presently done.

\section{Inclusive jet production}
\label{sec:jet}

\begin{figure}[htbp]
\vspace{135mm}
\includegraphics{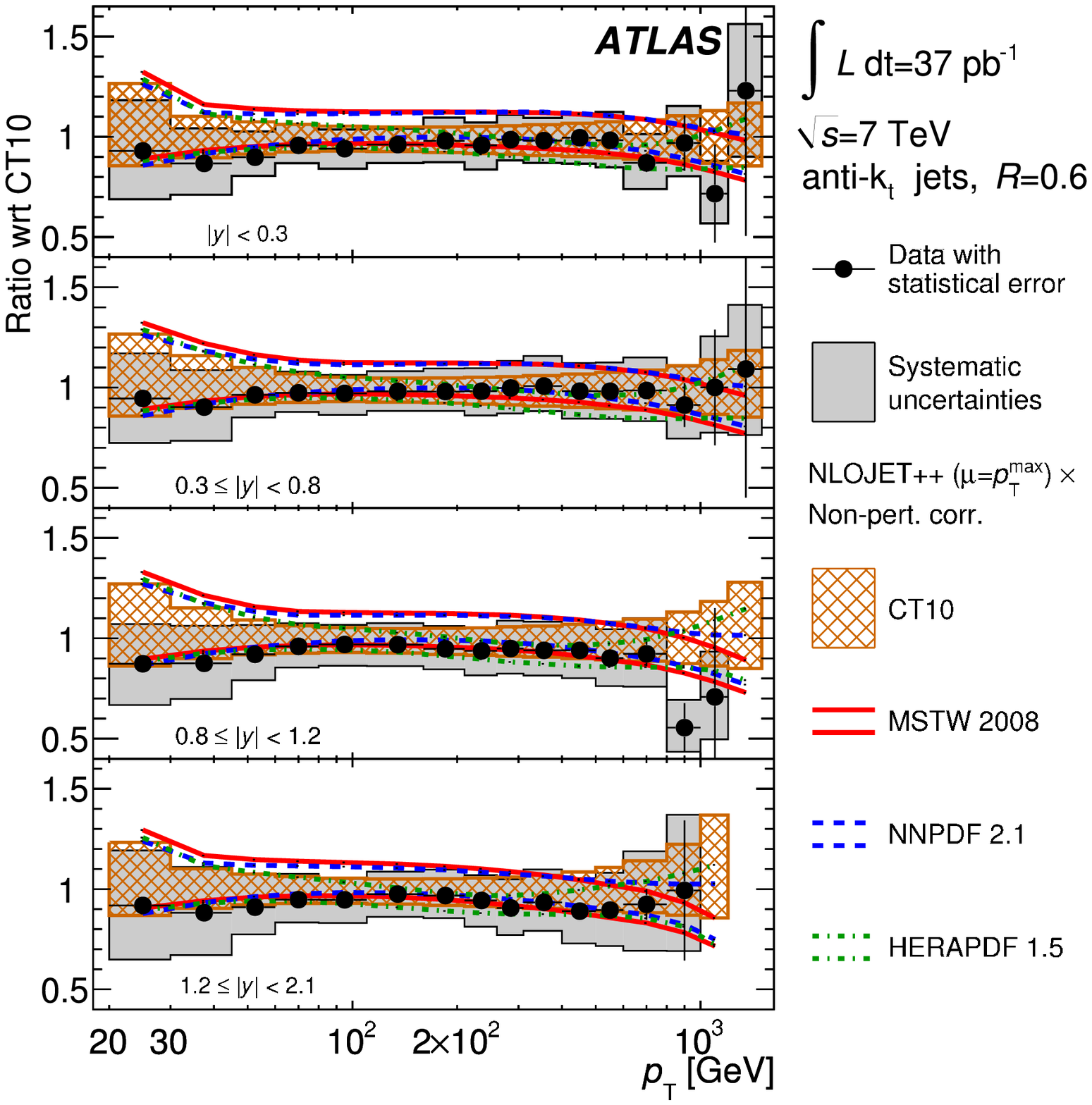}
\includegraphics{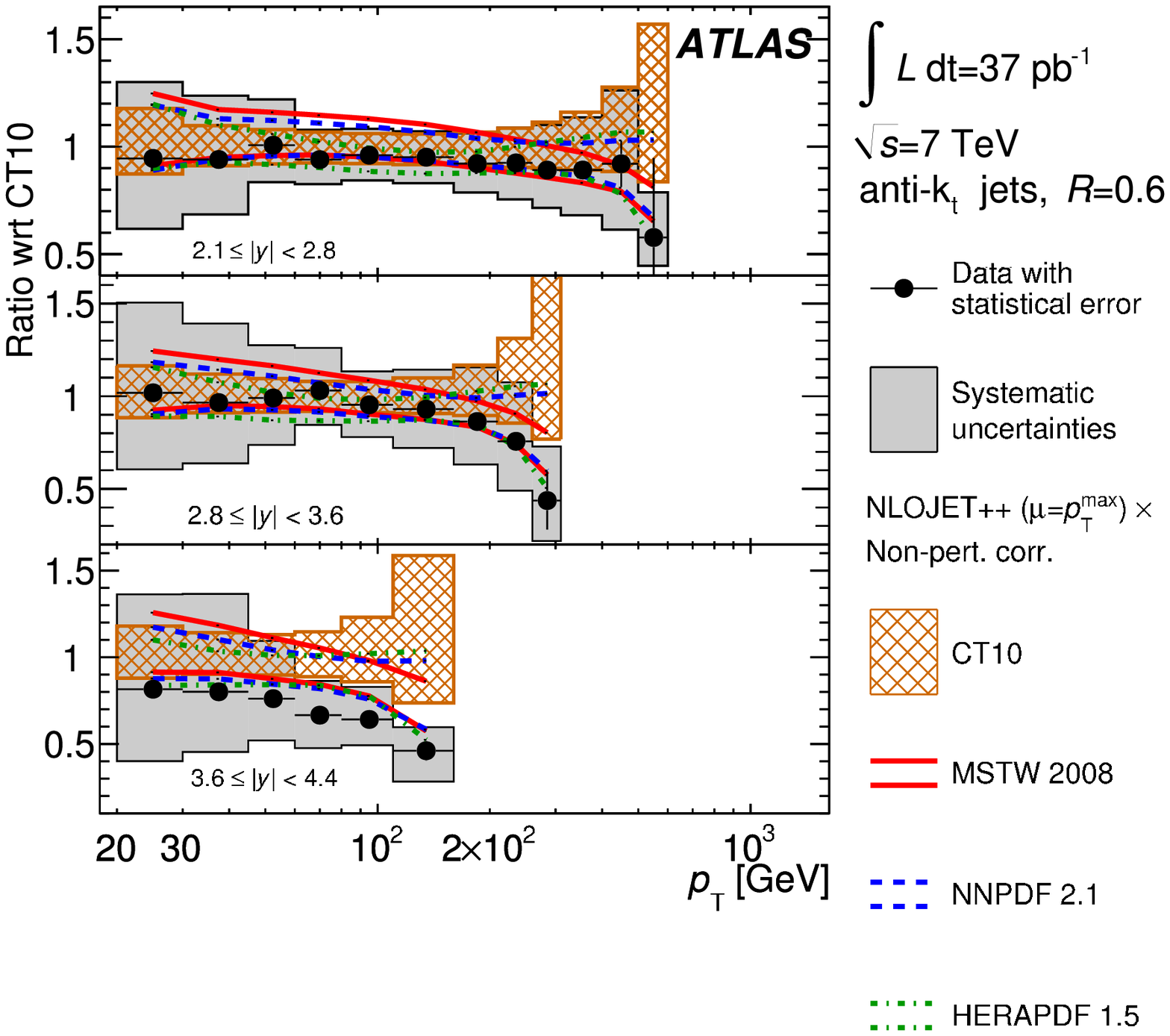}
\includegraphics{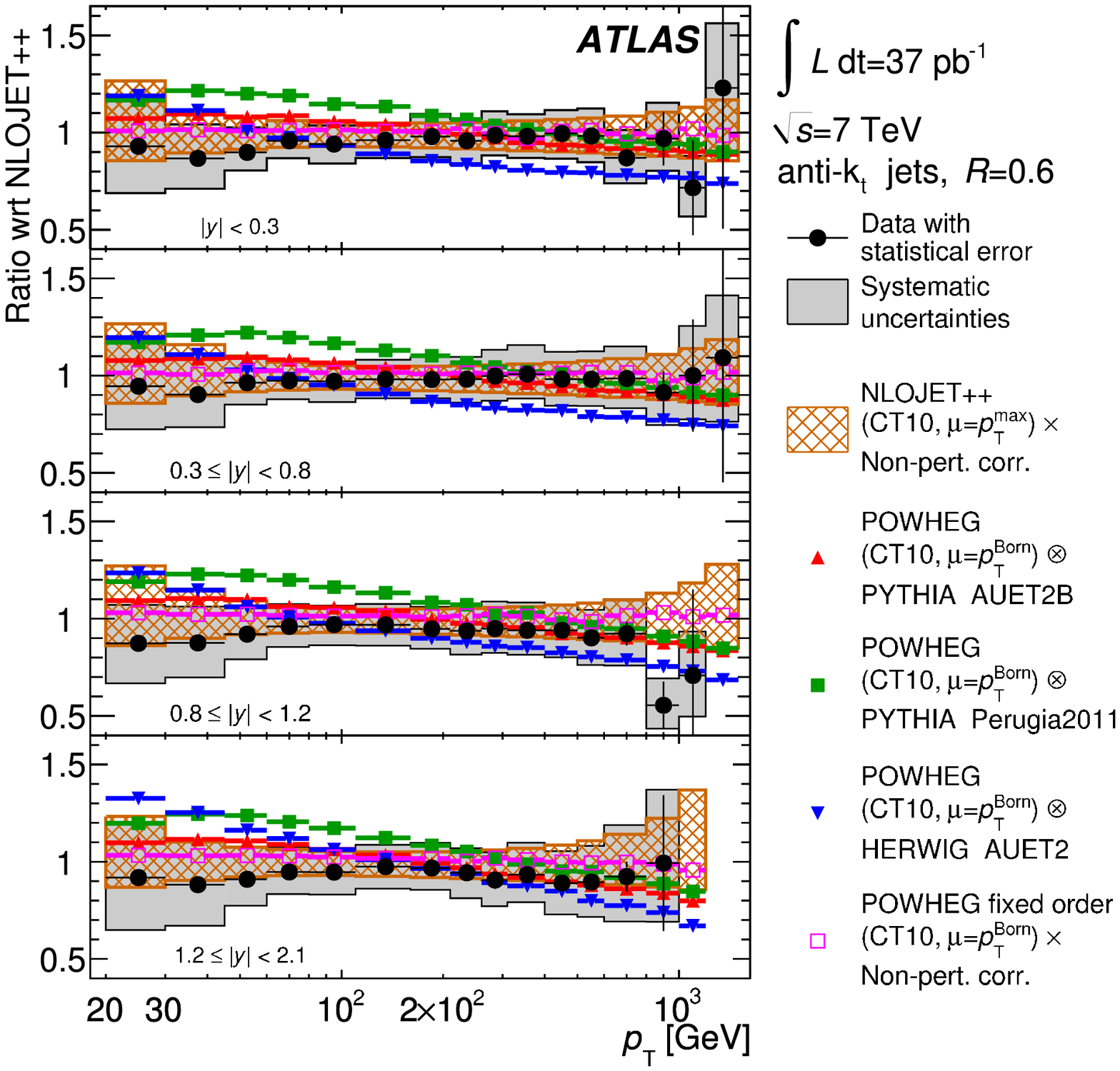}
\includegraphics{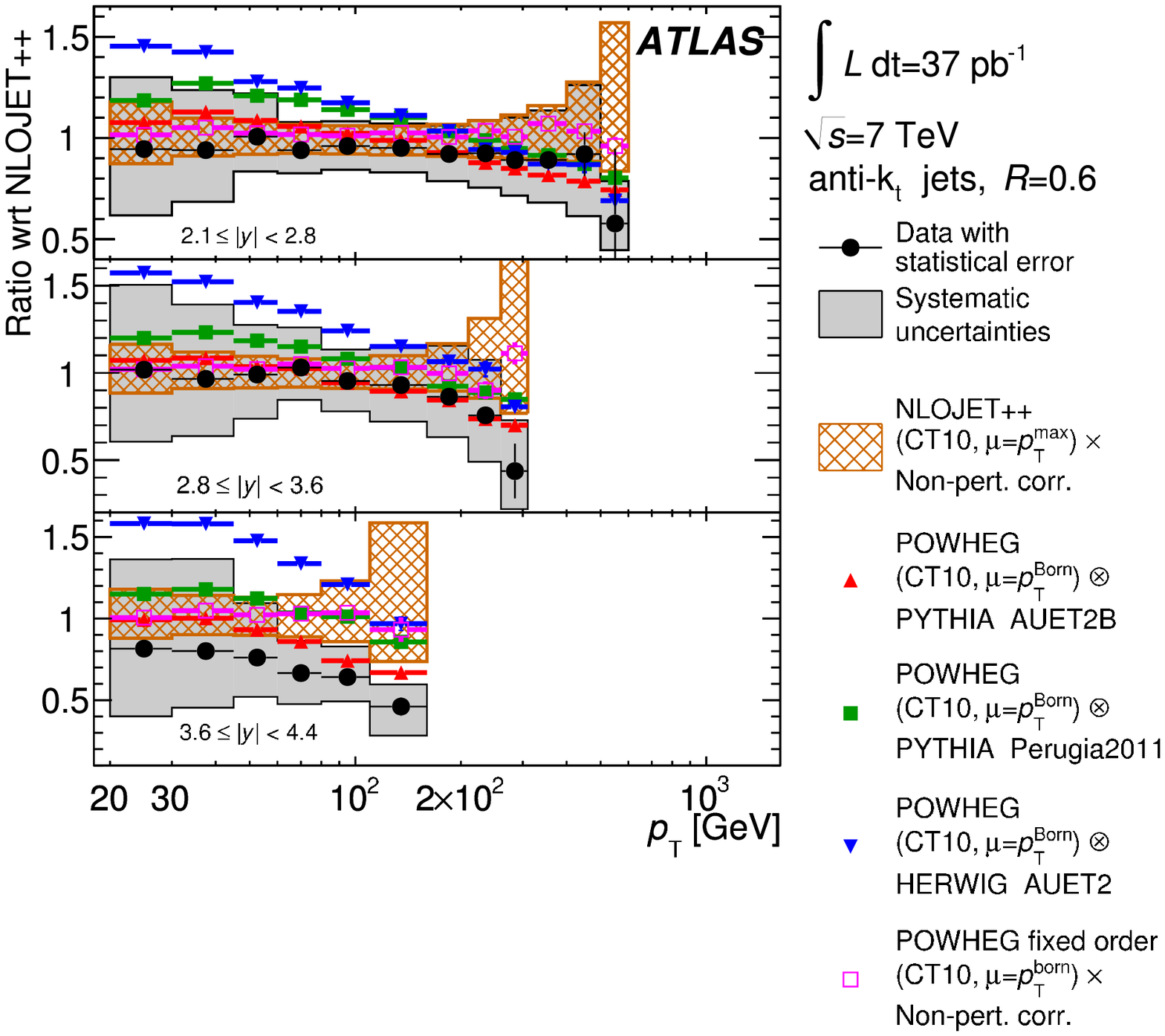}
\caption{\it  Inclusive jet spectra~\cite{atlas-1112} compared with   (top) NLO+NP results and 
(bottom)  NLO-matched shower results.} 
\label{fig:1112atlas-1}
\end{figure}

Measurements of inclusive jet production 
are   carried out 
at the LHC~\cite{atlas-1112,CMS:2011ab}  
 over  a     kinematic  range in transverse momentum 
and rapidity  much larger  than in any previous collider experiment~\cite{klaus-r}. 
Baseline comparisons  
 with Standard Model theoretical predictions   rely 
 either  on 
next-to-leading-order (NLO) QCD calculations, supplemented with 
nonperturbative  (NP) corrections~\cite{atlas-1112,CMS:2011ab} estimated from 
Monte Carlo  event generators,  or on NLO-matched parton shower event 
generators~\cite{ma,hoeche11}.  The 
upper panels 
 in  Fig.~\ref{fig:1112atlas-1}~\cite{atlas-1112} report the first kind of comparison, showing  
 that the NLO 
calculation agrees with data at central rapidities, while increasing deviations are 
seen with increasing rapidity at large transverse momentum $p_T$~\cite{atlas-1112}.   
The question  arises  of whether such    behavior is associated with 
 higher-order   perturbative contributions or with nonperturbative components 
 of the cross section.

The lower panels 
 in  Fig.~\ref{fig:1112atlas-1}~\cite{atlas-1112} 
show the second kind of comparison 
 based on  \powheg\  calculations~\cite{alioli},    
in which  NLO matrix elements are  matched with parton showers~\cite{herwref,pythref}.   
This improves the description of data, 
indicating that 
higher-order radiative  contributions  taken into account via 
parton showers are  
numerically  important.  At the same time, the results  show large differences between 
\powheg\  calculations interfaced with different shower generators, 
\herwig~\cite{herwref} and \pythia~\cite{pythref}, in the forward rapidity region.  
This  region is  sensitive to the details of parton showering  corrections. 

These observations make it  apparent that 
QCD contributions beyond NLO are essential for the 
 understanding of  LHC  jet data. 
In the following subsection we discuss showering and 
 nonperturbative  effects.  

\subsection{Nonperturbative and showering corrections}
\label{sec:np-sho}

 Using leading-order  Monte Carlo (LO-MC) generators~\cite{herwref,pythref}, the 
nonperturbative 
 correction factors  are schematically obtained  in~\cite{atlas-1112,CMS:2011ab} 
as 
\begin{equation} 
\label{npK1} 
K^{NP}_{0} =  { {  N_{LO-MC}^{(ps+mpi+had)} } /  {  N_{LO-MC}^{(ps)} }}  \;\; ,
\end{equation} 
where $(ps+mpi+had)$  and   $(ps)$ mean 
respectively  a simulation including   
parton showers, multiparton interactions and hadronization,  
and   a simulation including  only 
 parton showers     in addition to the LO hard process. 

While this  is a natural way to 
estimate NP   corrections from LO+PS event generators,  it is noted in~\cite{sampao} that
 when these  corrections  are combined   
 with NLO parton-level results a potential 
   inconsistency arises  because the 
    radiative correction  from the first  gluon emission   is treated  at different levels  of 
    accuracy     in the two parts of the  calculation.    To avoid this, 
Ref.~\cite{sampao}   proposes a method 
which  uses  
NLO Monte Carlo (NLO-MC) generators to determine the correction. 
In this case one can consistently  assign correction factors 
to be applied to  NLO calculations.  This  method  allows one to 
study  separately 
 correction factors to the  fixed-order calculation  due to  parton showering  
effects. 
To do this,  Ref.~\cite{sampao}    introduces  
  the correction factors $K^{NP}$  and  $K^{PS}$ as 
\begin{equation} 
\label{npK2} 
K^{NP} =  { {  N_{NLO-MC}^{(ps+mpi+had)} } /  {  N_{NLO-MC}^{(ps)} }}   \;\;   , 
\end{equation} 
\begin{equation} 
\label{npK3} 
K^{PS} =  { {  N_{NLO-MC}^{(ps)} } /  {  N_{NLO-MC}^{(0)} }}   \;\;   ,  
\end{equation}
where    the  denominator  in Eq.~(\ref{npK3})   is  defined  by switching off 
all components beyond NLO in the Monte Carlo simulation.

\begin{figure}[htbp]
\begin{center}
\includegraphics[scale=.24]{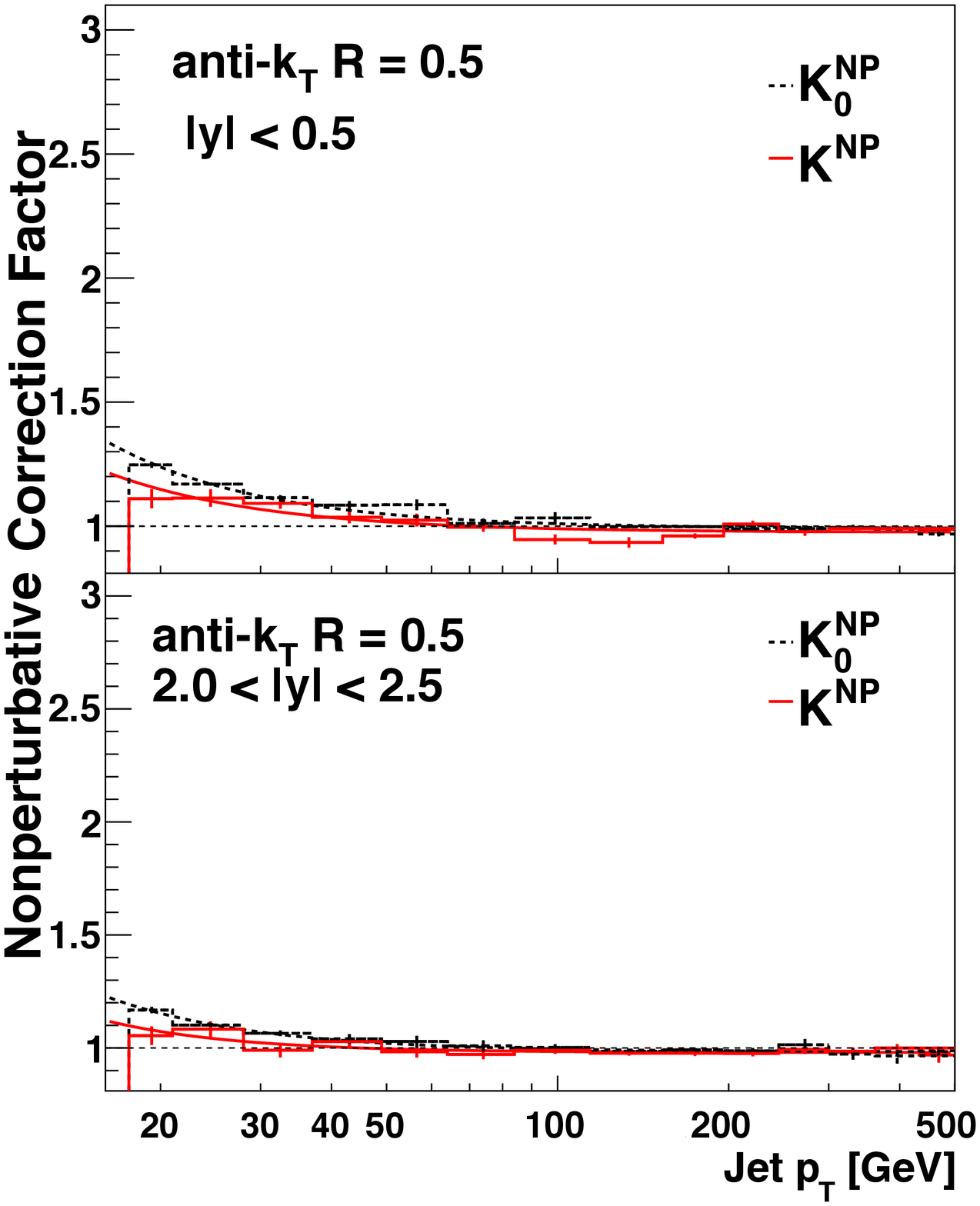}
\includegraphics[scale=.24]{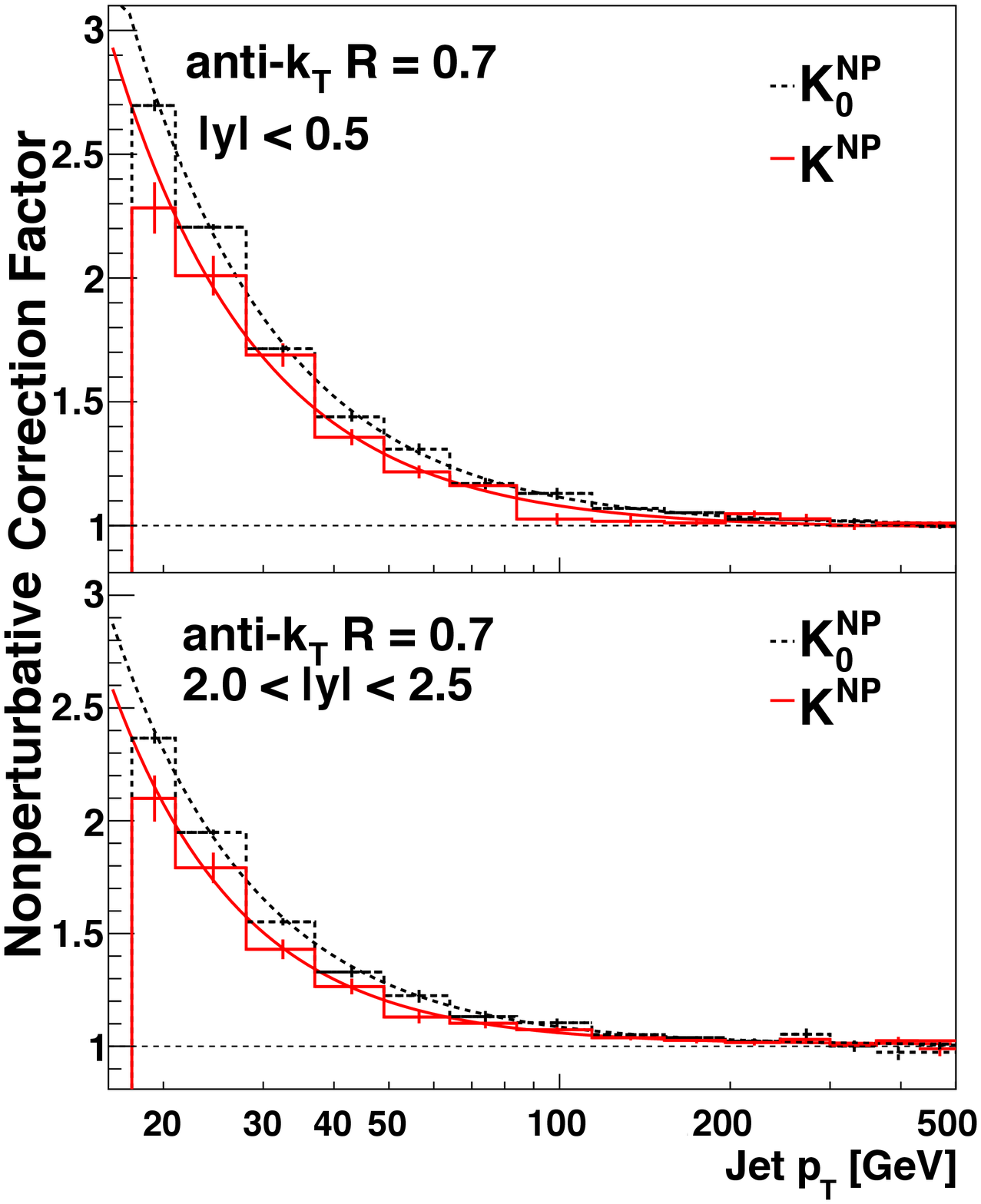}
\caption{\it The NP  correction factors to jet transverse 
momentum distributions  obtained from   Eq.~(\ref{npK1}) and 
 Eq.~(\ref{npK2}),  using 
 \protect\pythia\  and \protect\powheg\  respectively, 
 for $|y|<0.5$ and $2 < |y|< 2.5$.  
Left: $R=0.5$; Right:  $R=0.7$.~\cite{sampao}}  
\label{fig:np1}
\end{center}
\end{figure}

\begin{figure}[htbp]
\begin{center}
\includegraphics[scale=.24]{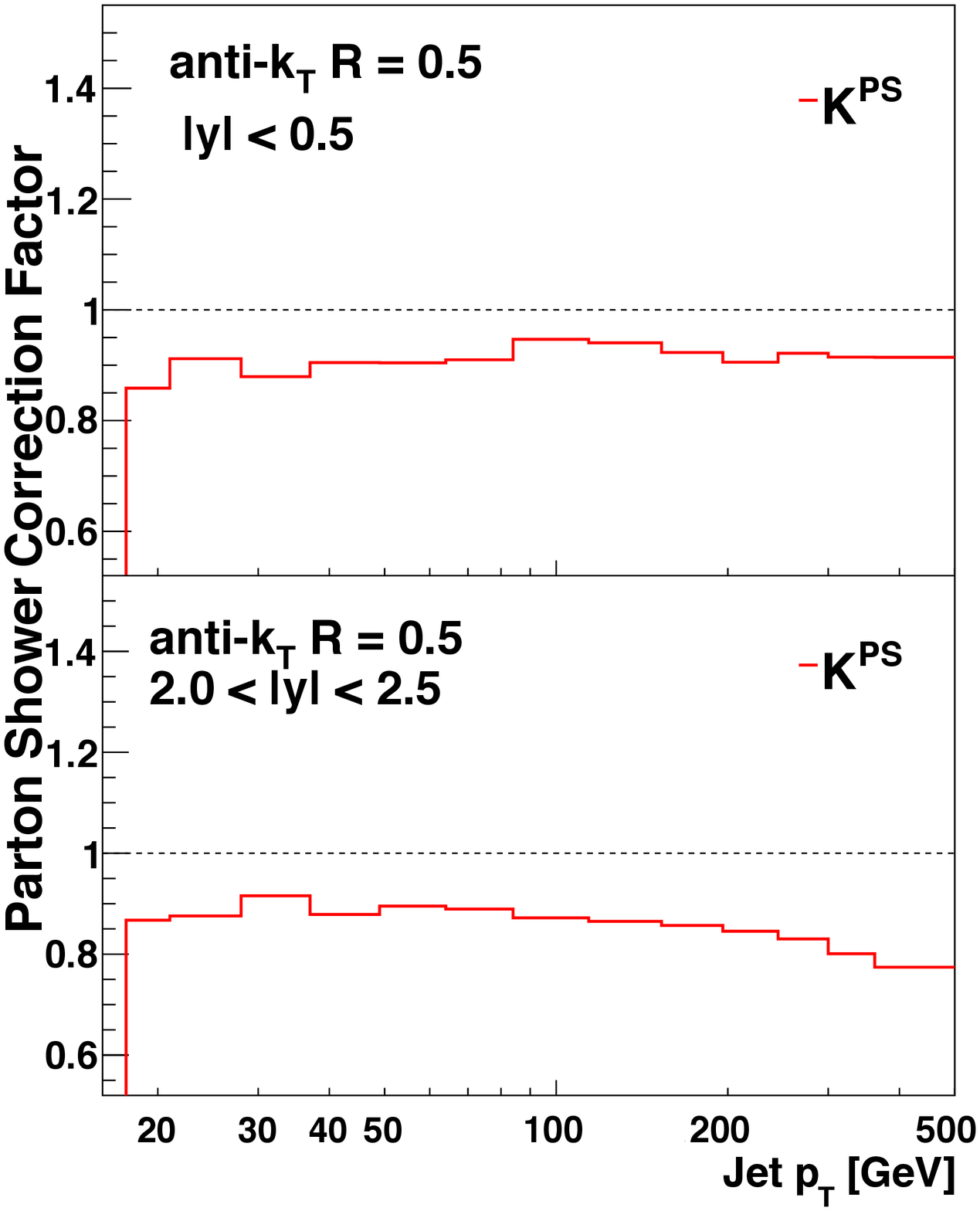}
\includegraphics[scale=.24]{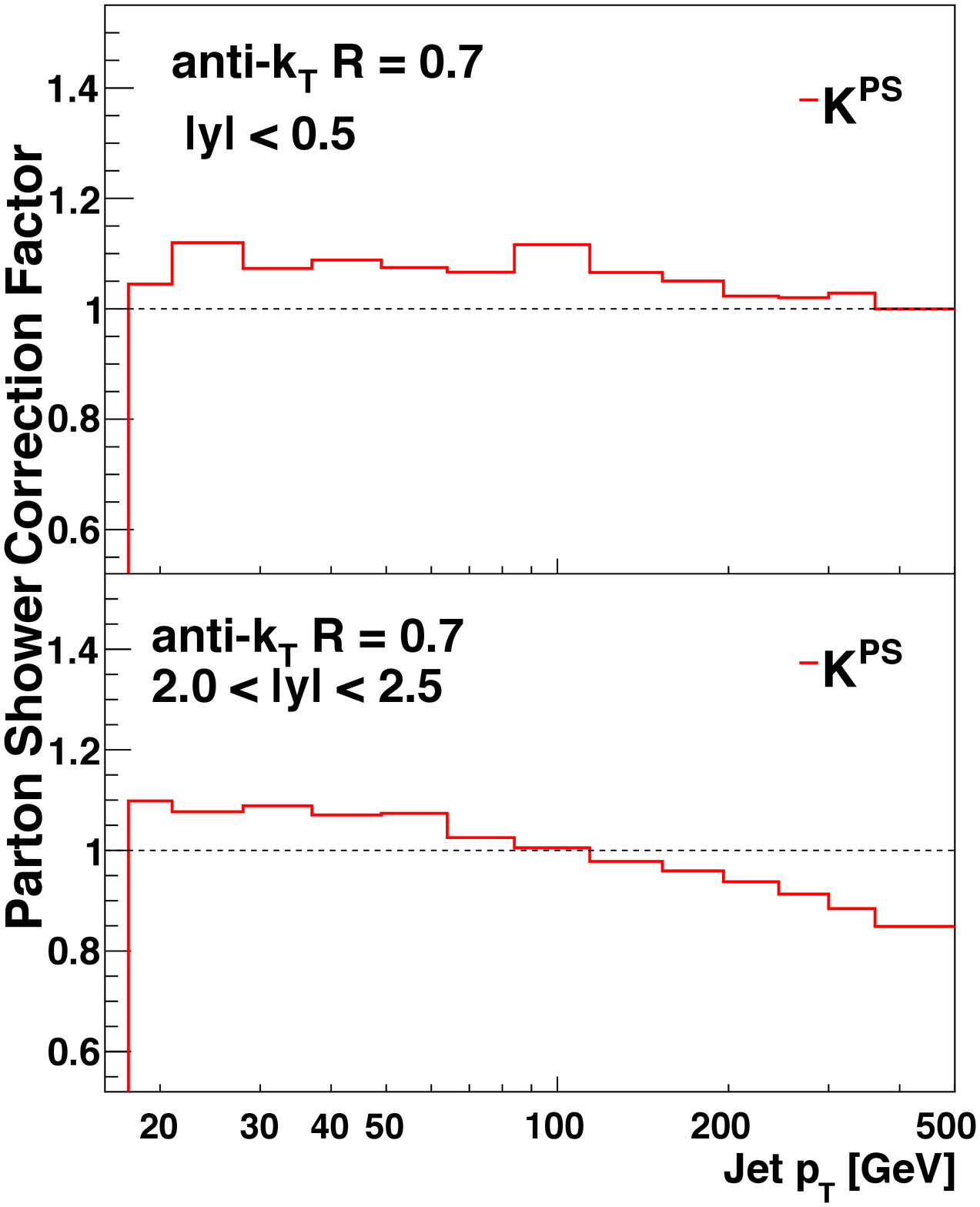}
\caption{\it The  parton shower
 correction factor to jet transverse momentum  distributions, obtained from  
 Eq.~(\ref{npK3})  using 
 \protect\powheg\   
   for $|y|<0.5$ and $2 < |y|< 2.5$.
Left: $R=0.5$; Right:  $R=0.7$.~\cite{sampao}}  
\label{fig:np2}
\end{center}
\end{figure}

The factor $ K^{NP} $ in  Eq.~(\ref{npK2}) 
differs from $ K^{NP}_{0}  $ because of the different definition 
of the hard process. In particular the multi-parton interaction p$_T$ cut-off scale is 
different in the LO and NLO cases. 
Numerical results are shown in  Fig.~\ref{fig:np1}. 
The factor $ K^{PS} $ in  Eq.~(\ref{npK3}), on the other hand,   
  is new. It singles out 
contributions due to parton showering and has not been considered before. 
Unlike the NP correction,  it gives finite effects also at large p$_T$. 
The results in 
 Fig.~\ref{fig:np2}  show in particular  that 
 this    correction is not merely a  rescaling but 
  is $y$ and p$_T$ dependent, especially when rapidity is 
non-central.

\begin{figure}[htbp]
\begin{center}
\includegraphics[scale=.5]{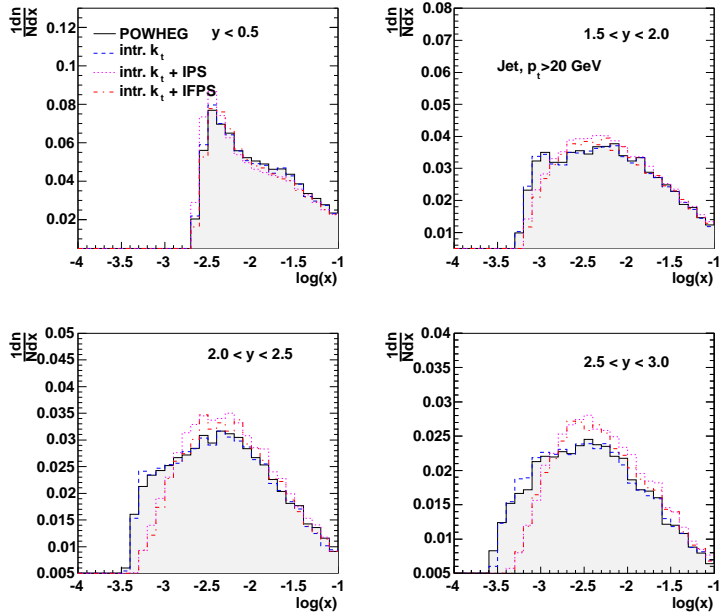}
\caption{\it Distributions in the parton longitudinal momentum fraction $x$  before (POWHEG) and after parton showering (POWHEG+PS),  
for  inclusive jet  production  at  different rapidities for jets with $p_T> 18 $ GeV obtained 
by  the anti-kt jet  algorithm~\cite{antiktalgo} with $R=0.5$.
Shown is the effect of intrinsic $k_t$, initial (IPS) and initial+final state (IFPS) parton shower.~\cite{sampao}}  
\label{fig:fig1}
\end{center}
\end{figure}

The result  in  Fig.~\ref{fig:np2} comes from initial-state and final-state showers. 
 These are   interleaved   so   that the 
      combined effect is nontrivial and cannot be obtained 
     by simply adding the two~\cite{sampao}.   
   In general the effect from parton shower is largest at large $|y|$, where  the 
initial-state parton shower is mainly contributing 
at low $p_T$, while the final-state parton shower is contributing significantly 
 over the whole $p_T$ range.   It is observed in~\cite{sampao} that 
  the main initial-state  showering effect comes 
from  kinematical  shifts  in   longitudinal momentum   distributions~\cite{coki-1209},    
due to combining collinearity approximations   with  the Monte Carlo 
implementation of  energy-momentum conservation constraints. 
The effect of the kinematic shifts is  illustrated in  Fig.~\ref{fig:fig1}~\cite{sampao}, showing 
 the distribution in the parton longitudinal momentum fraction $x$ 
before  parton showering 
and after parton  showering.  We see that the longitudinal shift is negligible 
 for  central rapidities but becomes 
significant for $ y >1.5$.  

 In summary, 
the nonperturbative correction factor $K^{NP}$ introduced 
from NLO-MC  in Eq.~(\ref{npK2})   
gives non-negligible differences 
compared to the LO-MC contribution~\cite{atlas-1112,CMS:2011ab} 
at low to intermediate jet $p_T$, while 
  the showering  correction factor  $K^{PS}$  of  Eq.~(\ref{npK3})   
  gives  significant effects  over the whole $p_T$  range  and is largest at 
  large  jet  rapidities $y$.   
 Because of this     $y$ and $p_T$ dependence,   
    taking properly into account    NP and showering correction factors     
  changes  the  shape  of jet distributions, and   may  thus  influence   the comparison 
  of theory predictions with experimental data. 
We anticipate  in particular 
that   taking  account  of  the  showering correction factor    
  will    be     relevant     in    fits  for 
parton distribution functions   using  inclusive jet data.

\section{Forward jets}
\label{sec:forw}

A   program of jet physics in the forward region 
can be carried out for the first time at the LHC~\cite{fh-aoste}, by    
    exploiting   the large phase space opening  up at  high  center-of-mass energies,  
and  the unprecedented reach in rapidity 
 of  the experimental instrumentation~\cite{ajaltouni}. 
First forward jet measurements have been performed by 
 LHC experiments~\cite{cms1202,atlasjetveto,dijetratios}. 

The evaluation of QCD theoretical predictions  for forward jets is made complex by 
 high-$p_T$ production  occurring in  a region characterized by multiple 
hard   scales, possibly widely disparate from each other~\cite{fh-aoste}.    
In addition to the  parton-shower kinematic effects  discussed in Sec.~2, 
   when jets are measured 
at large separations in rapidity   dynamical contributions   from  soft  multi-gluon  
radiation~\cite{muenav,cch,jhep09} set in,  calling for perturbative resummations of  
 large-rapidity logarithms to all orders in $\alpha_s$.  This motivates current 
studies based on the BFKL equation~\cite{samuel,jeppe,martin-uam,arcava}. 
Furthermore, with increasing  rapidities  
    the nonperturbative  parton  distributions  are probed  in highly asymmetric 
kinematic regions   near the boundaries  $ x \to 0$ and $x \to 1$ of partonic  phase 
space~\cite{fh-aoste,jhep09}.   
This  in turn   implies  that contributions from    multiple parton collisions~\cite{pavtrel82,sjozijl}  
are potentially enhanced~\cite{bartal,yuri-talk,markus-talk}.    
Effects of  multiple scatterings  may be  studied   
in di-jet as well as di-hadron spectra~\cite{di-hadr-raju},  
in pp collisions and  in collisions of light~\cite{trel12} and 
 heavy~\cite{albac-etal,sapeta,marquet-xiao} nuclei.

 Forward jets  enter  
the LHC physics program   in  an  essential 
way both for   new particle discovery 
processes (e.g.,    Higgs 
searches~\cite{vvfusion}   in    vector boson fusion channels,  jet studies in  decays of    
highly boosted heavy states~\cite{alth-boost}) and for  new aspects of 
standard model physics  (e.g.,  small-$x$ QCD  and its   
   interplay with cosmic ray physics~\cite{cosmic-lhc,grothe}, 
studies of   high-density  parton 
 matter~\cite{iancu12,denterria}).    
The first forward jet measurements~\cite{cms1202,atlasjetveto,dijetratios}     
show  that, 
while inclusive forward jet spectra are roughly in agreement 
with predictions from different  Monte Carlo   simulations, 
detailed aspects of production rates  
 and   correlations      
are not  well    understood yet.  Recent phenomenological analyses 
are carried out in~\cite{powheg-hej,marquet-royon}.

\begin{figure}[htbp]
\vspace{65mm}
\includegraphics{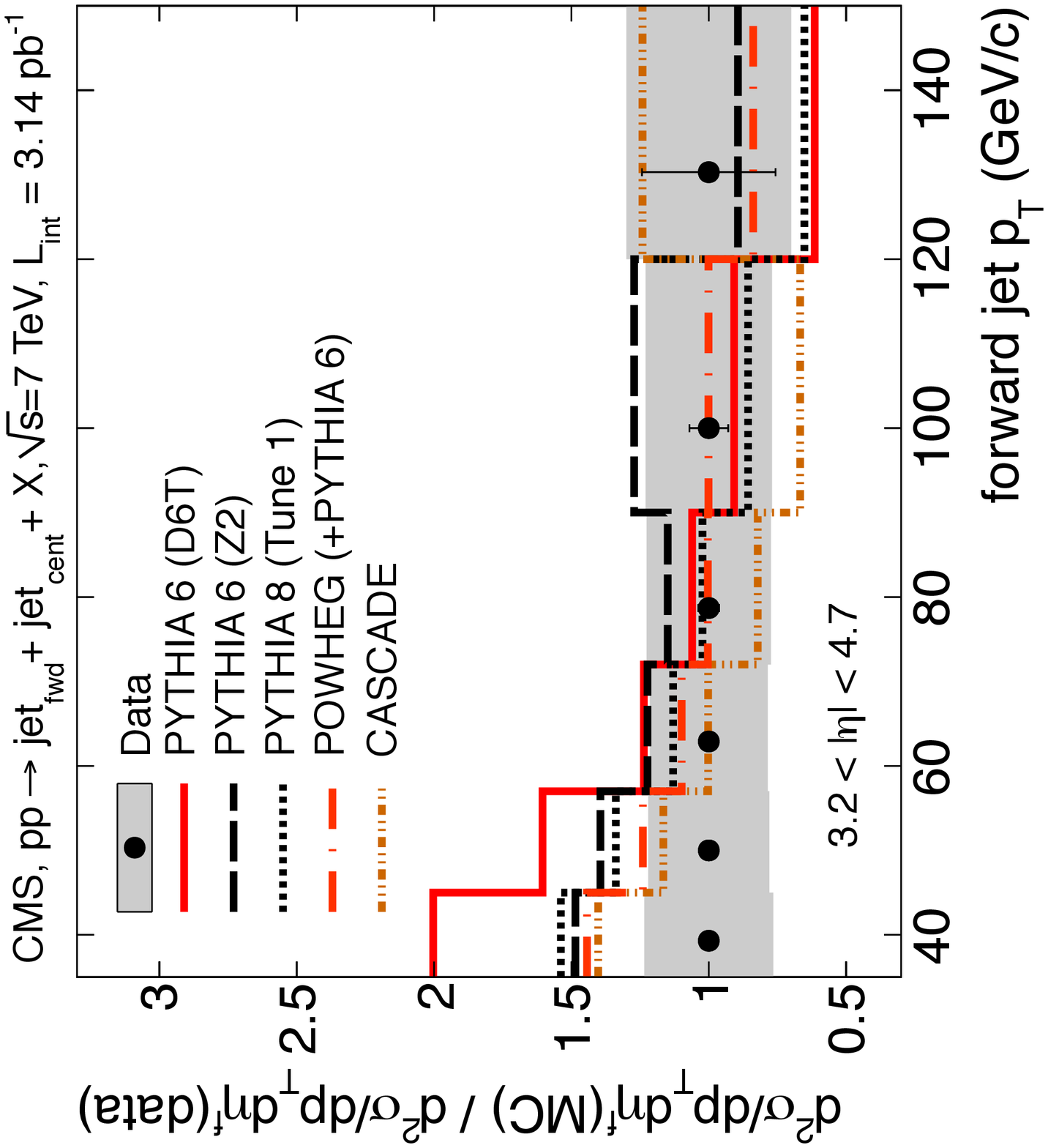}
\includegraphics{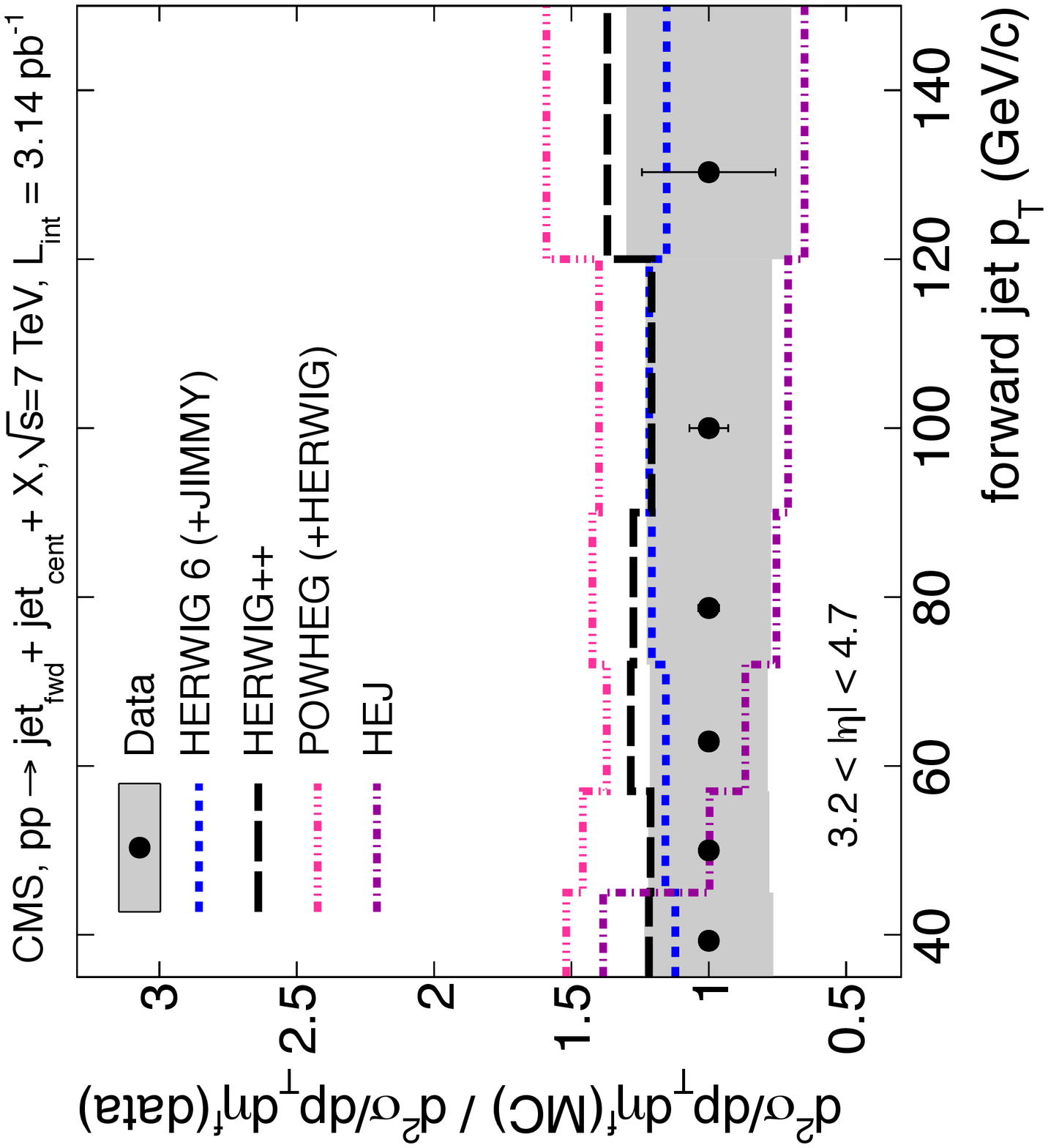}
\caption{\it   Ratio theory/data~\cite{cms1202} for dijet events with a central  
and a forward jet as a function of the forward jet transverse momentum.} 
\label{fig:12020704-fig2}
\end{figure} 

An example 
is given by the di-jet 
observables~\cite{epr1012} associated with events 
containing a forward and a central jet. 
Experimental measurements and Monte Carlo comparisons are shown 
in Fig.~\ref{fig:12020704-fig2}~\cite{cms1202}. The results  indicate  that    
none of the Monte Carlo generators describes  the data  well in 
all regions;   
 in particular   NLO-matched calculations  from \powheg\ give 
large differences  in the forward jet  p$_T$ distribution 
when combined with different parton showers,   see 
\powheg+\herwig\  vs.  \powheg+\pythia.

\begin{figure}[htbp]
\vspace{60mm}
\includegraphics{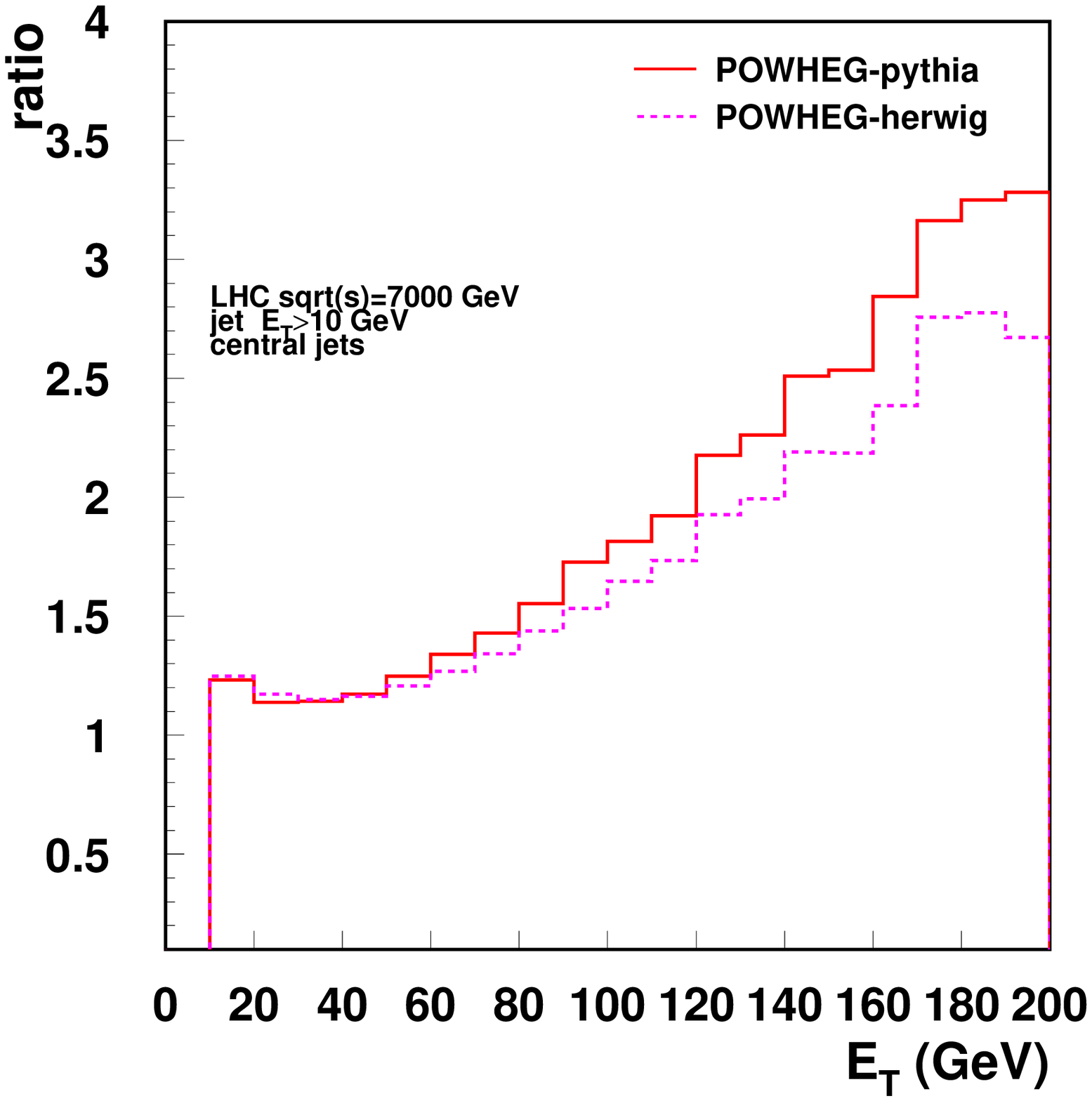}
\includegraphics{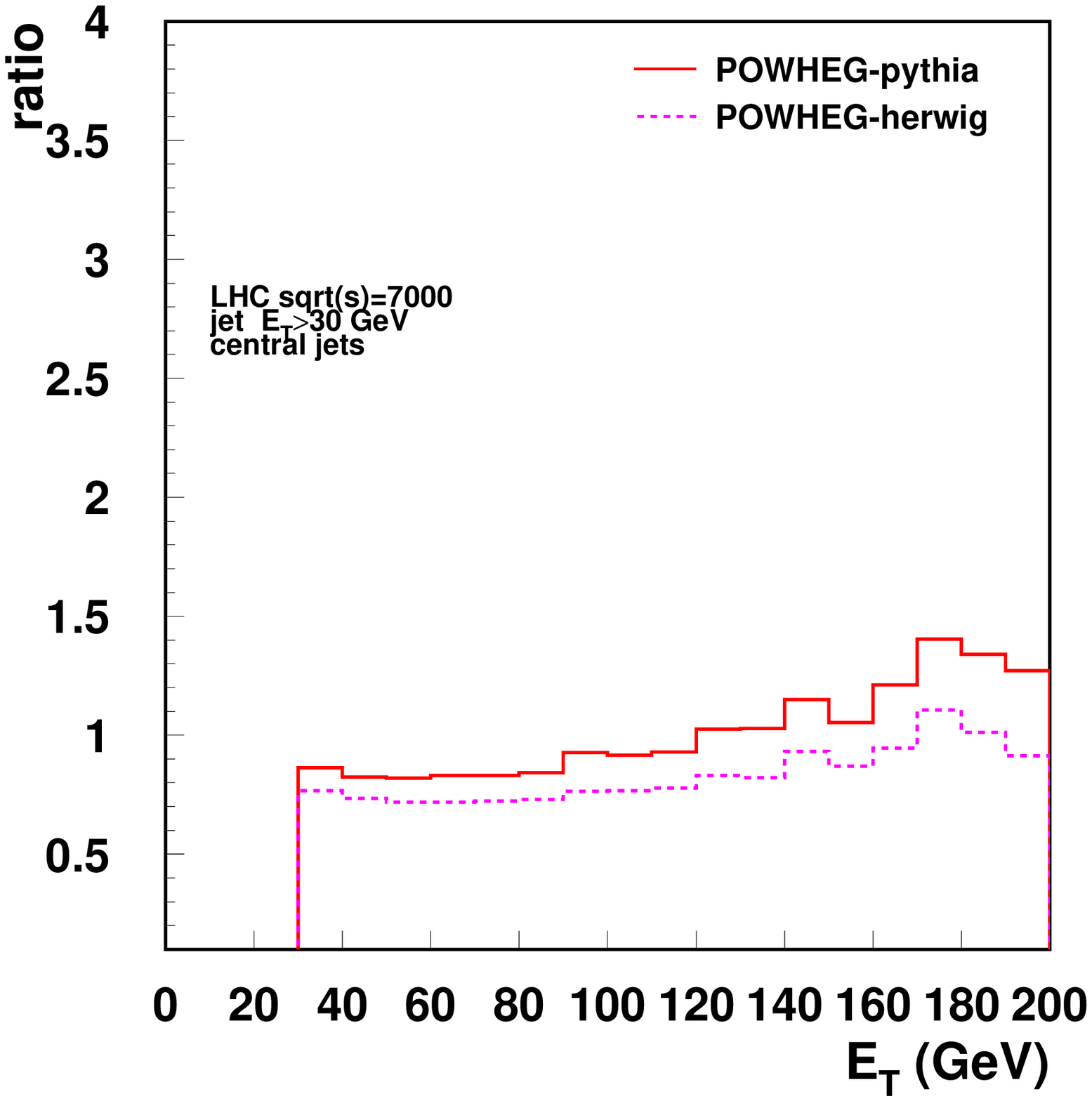}
\caption{\it Ratio of  NLO+shower to no-shower results  
for dijet events with a central  
and a forward jet as a function of the central 
 jet transverse momentum~\cite{1206shower}. 
(left) $E_T > 10$~GeV; (right) $E_T>\!30$~GeV. } 
\label{fig:powhegshowers}
\end{figure}

In~\cite{epr1012,1206shower} this behavior is investigated by  studying
 $\Delta R $ jet distributions designed to provide a 
measure, in azimuth and rapidity space,  
of the extent to which 
 jets are dominated by hard partons in the matrix element  or 
originate from showering. Large contribution to  jets 
from  showering are found~\cite{epr1012} 
in the case of asymmetric parton kinematics,  i.e, when one of the 
 initial-state showers goes down to small $x$. 
Ref.~\cite{1206shower} furthers this study 
   by considering the central jet transverse energy spectrum, in 
di-jet events  with a central and a forward jet,  
using the NLO event 
generator \powheg\  matched with   parton showers   
   \pythia\ and \herwig. 
Fig.~\ref{fig:powhegshowers}~\cite{1206shower}   shows results 
 for the two cases, 
normalized to the result obtained by  switching off parton showering.  The 
marked differences between the two cases  are consistent 
with the findings in~\cite{cms1202}, and with the  large  
contribution to  jets from  showering found in~\cite{epr1012}.  
In particular  in the forward-central events considered 
  high-rapidity  correlations  turn out  to  affect  the behavior  of 
jet distributions in the central region. 

\begin{figure}[htbp]
\vspace{65mm}
\includegraphics{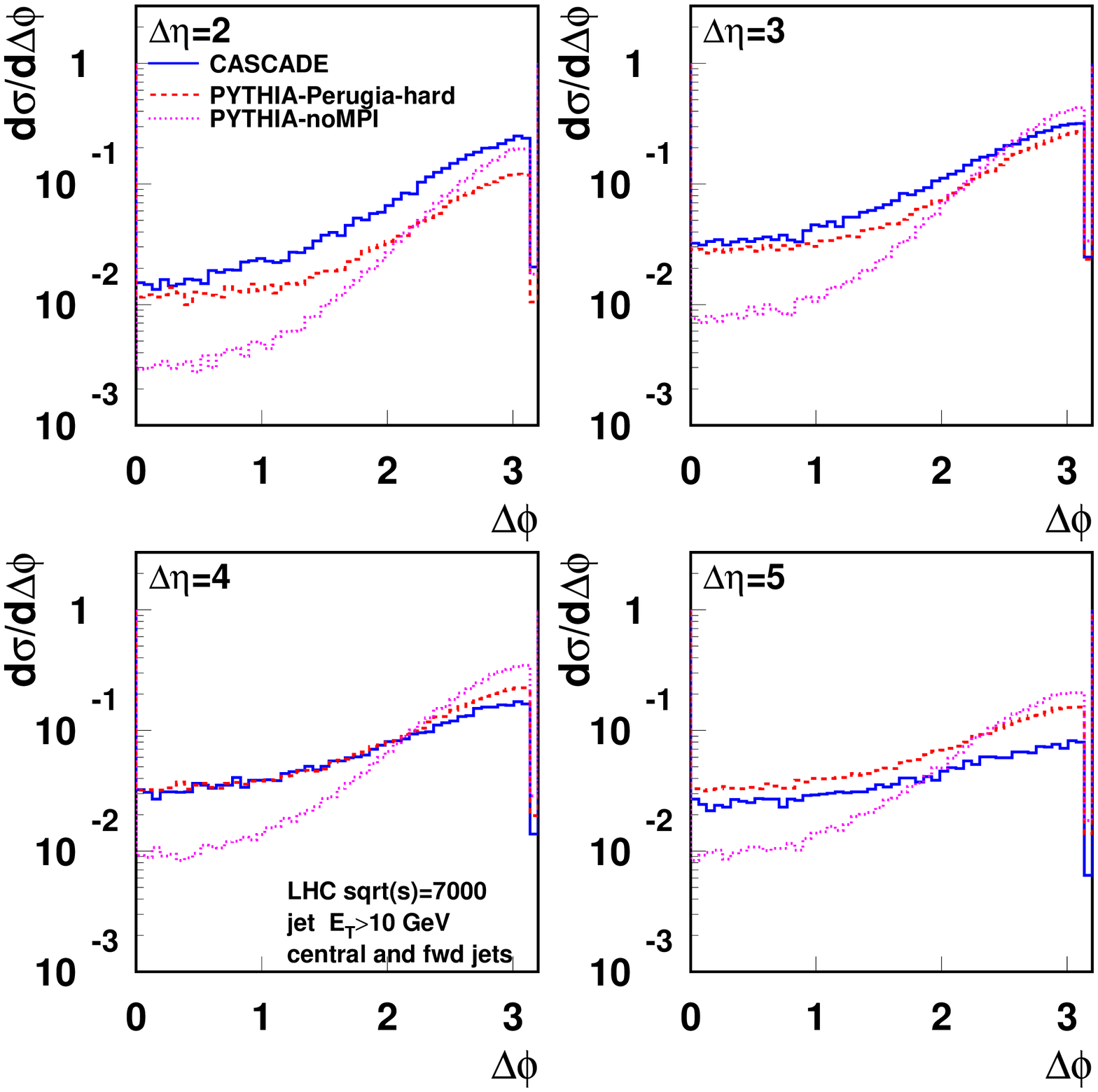}
\includegraphics{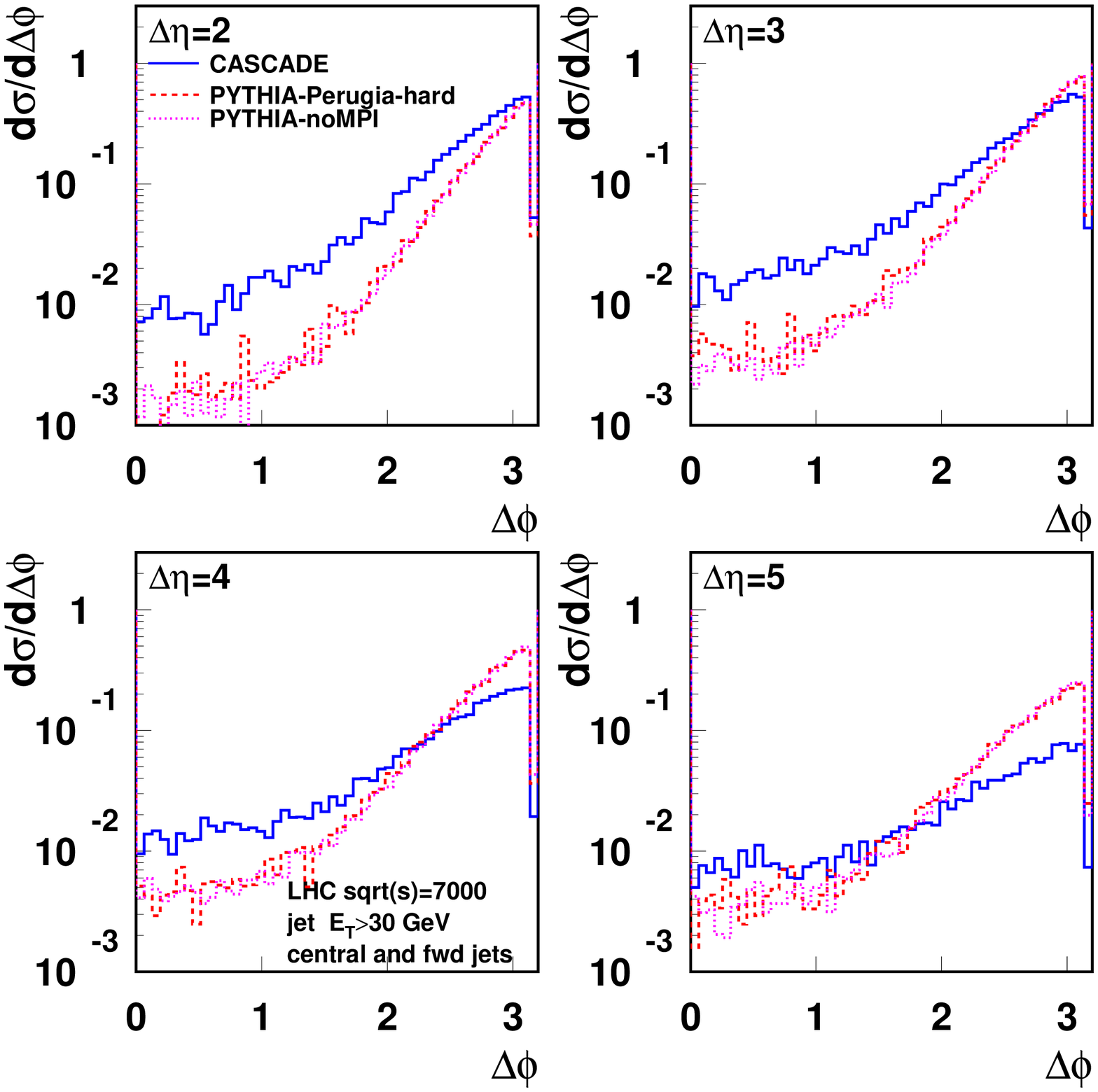}
\caption{\it Cross section versus  azimuthal distance  
$\Delta \phi$  between  central and  forward jet,  at  different  
rapidity  separations $\Delta \eta$,  for jets with transverse 
energy $E_T > 10$ GeV (left)  and 
$E_T > 30$ GeV (right)~\cite{epr1012}. } 
\label{fig:azsigma}
\end{figure}

A  classic test of  QCD high-energy resummation for  jets at large 
rapidity separations~\cite{muenav} 
 is given by the azimuthal decorrelation between  jets. 
Fig.~\ref{fig:azsigma}~\cite{epr1012}   shows   the  cross section 
 as a function of the 
azimuthal distance  $\Delta \phi$  between  central and  
forward jets reconstructed  with the   
   Siscone algorithm~\cite{fastjetpack} ($R =  0.4$), 
  for  different  rapidity separations. It shows results 
computed by 
\pythia\ Monte Carlo~\cite{pz_perugia}, with and 
without multi-parton interactions, and 
by \cascade\ Monte Carlo~\cite{cascade_docu}, which includes small-$x$ 
gluon coherence effects~\cite{coh-x}  in the initial-state shower. 
The main point is  that    
the decorrelation as a function of 
$\Delta\eta$ increases in \cascade\  as well as in \pythia, 
respectively as a result of  
finite-angle gluon  radiation in  single-chain parton shower  or as a result 
of multiple-chain collinear showers;  
but  while  in the low $E_T$  region (Fig.~\ref{fig:azsigma} (left)) 
this   is similar  
 between \cascade\ and \pythia\  with multiparton interactions   for $\Delta\eta < 4 $, 
in the higher $E_T$ region    
(Fig.~\ref{fig:azsigma} (right))
 the influence of multiparton interactions in \pythia\ is small and 
\cascade\ predicts everywhere  a larger decorrelation. 
 We will come back to this and   discuss 
  correlations  further  
in Sec.~5  from the point of view of energy flow  observables.

While  the specific 
results shown  in this section  refer to  forward-central jet correlations, 
it is    interesting  to consider 
extensions to the forward-backward kinematics. This will allow one  to 
address the large-$\Delta y$  di-jet data sets~\cite{atlasjetveto,dijetratios}, currently rather 
poorly understood;  to search 
for Mueller-Navelet effects~\cite{muenav};  to analyze 
 backgrounds in  Higgs boson studies~\cite{vvfusion} from  vector boson fusion 
channels.  In particular,     one may be able to extract information  on  
  Higgs properties  and couplings from jet kinematics~\cite{rohini}.  
In this case too     finite-angle radiative contributions to single-chain  showers,  
extending across  the whole  rapidity range, 
   affect the underlying  jet activity accompanying the Higgs~\cite{deak_etal_higgs}    
and  may  give competing effects to multiple-parton interactions.

\section{$B$-jets}   
\label{sec:heavy}

\begin{figure}[htbp]
\vspace{60mm}
\includegraphics{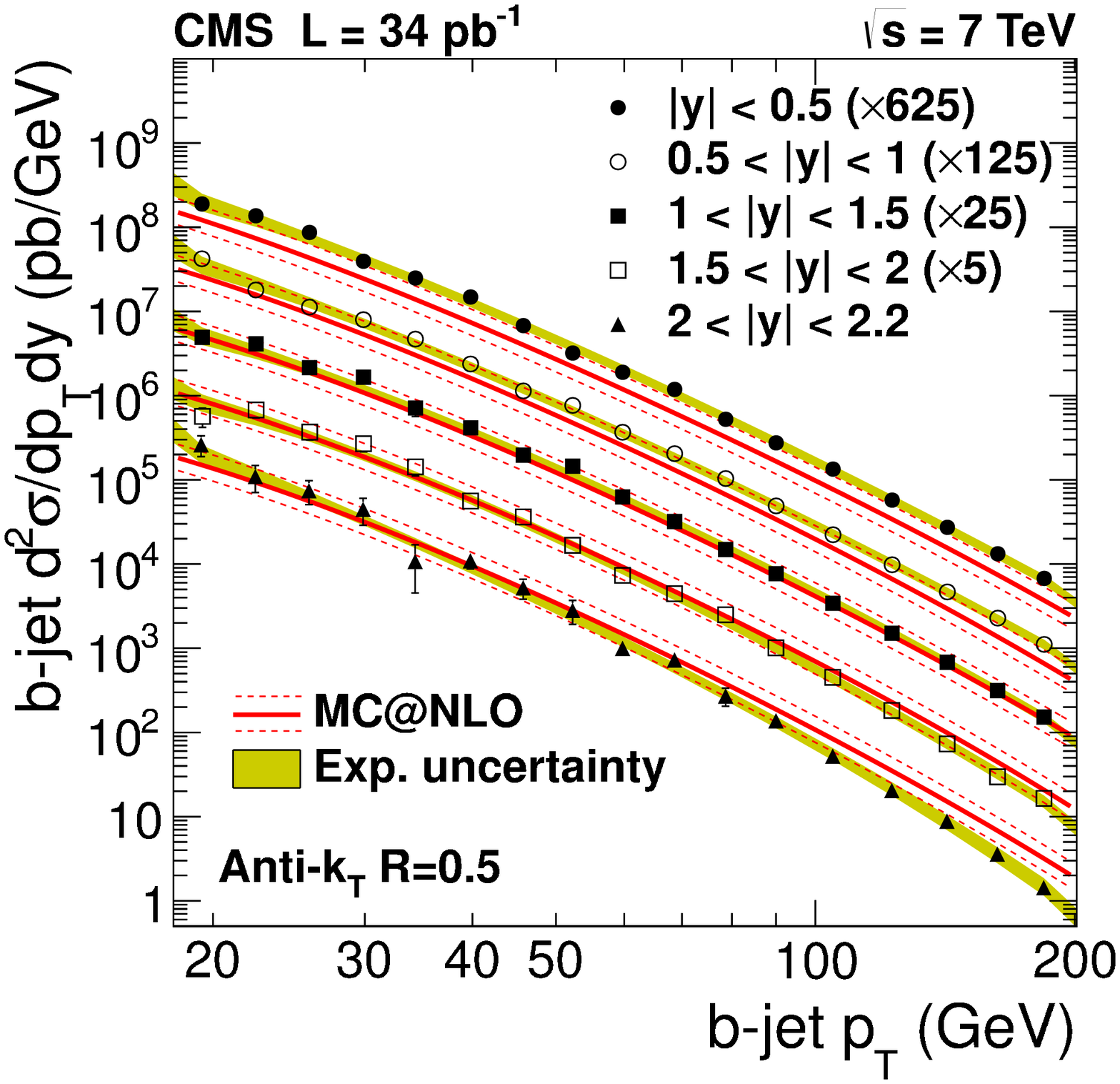}
\includegraphics{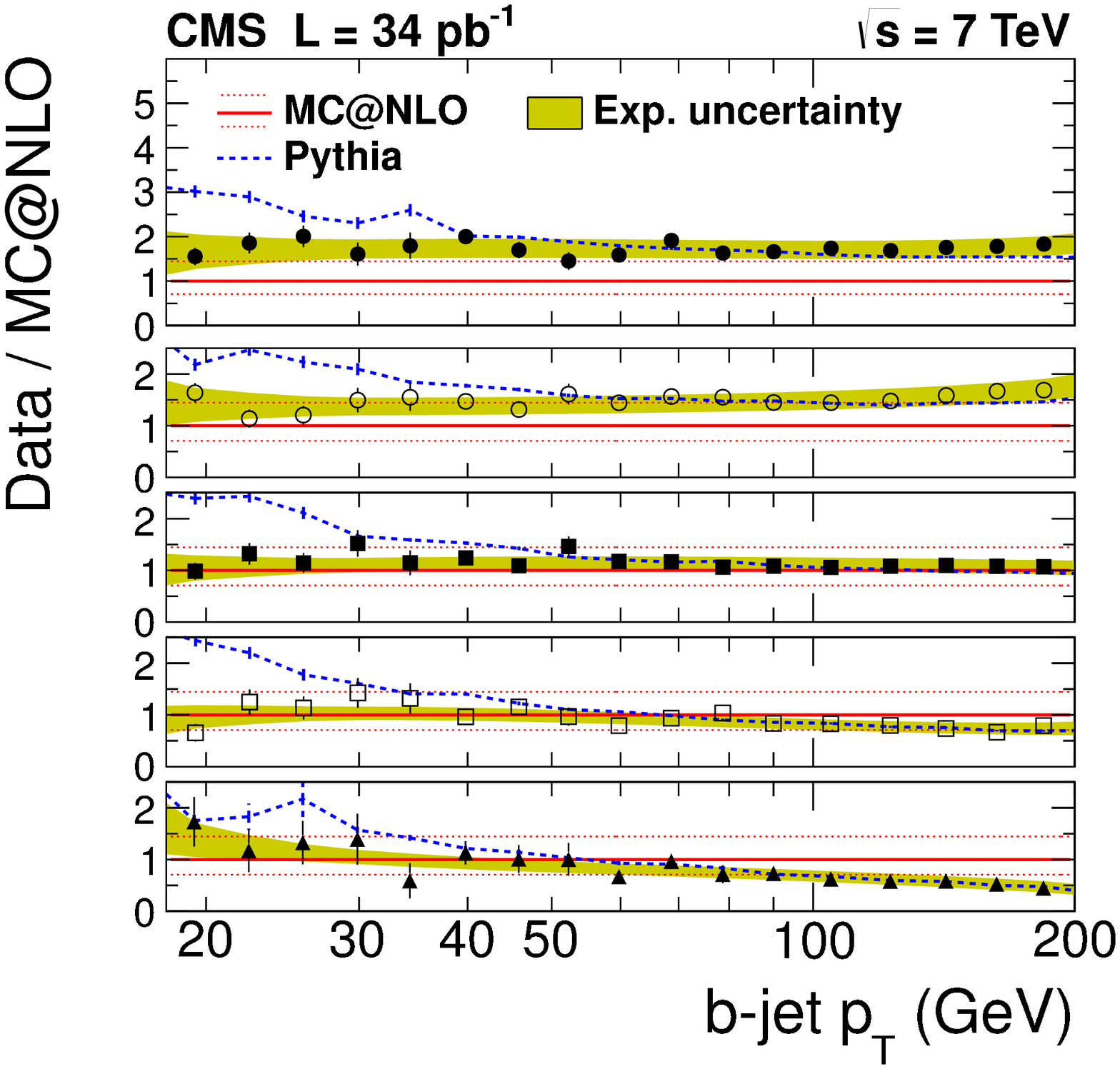}
\caption{\it  Inclusive $B$-jet spectra~\cite{cms1202-b}. } 
\label{fig:1202cms-b}
\end{figure}

Some  of the features observed for inclusive jets in Sec.~2 are also 
present in $b$-flavor  jets~\cite{cms1202-b,atlas1109-b}. 
Fig.~\ref{fig:1202cms-b}~\cite{cms1202-b} shows a comparison of 
the measured $B$-jet  transverse 
momentum spectra  with matched NLO-shower calculations using \mcatnlo~\cite{mc-nlo03}. 
The description of the data  is generally good at central rapidities, while  
at large $y$ and large  $p_T$  the Monte Carlo is above the data. 
 Similar  behavior is shown by  comparisons  with  
\powheg~\cite{frix-pow} in~\cite{atlas1109-b}.

\begin{figure}[htbp]
\begin{center}
\includegraphics[scale=.6]{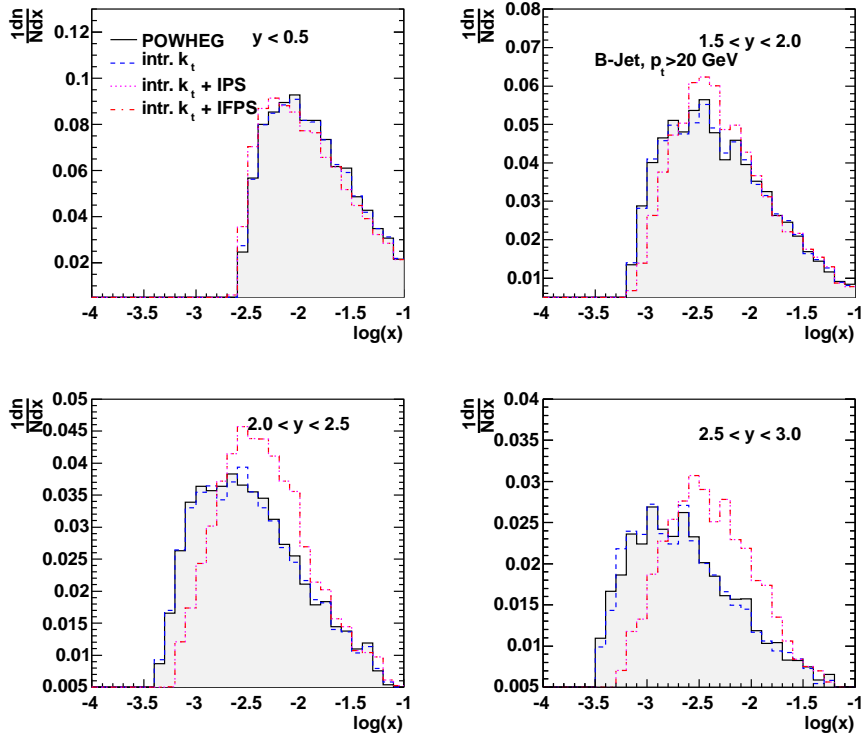}
\caption{\it   Production of $B$-jets: distribution in the parton 
longitudinal  momentum fraction $x$,  
before and after parton  showering,  for different rapidity regions.
Shown is the effect of intrinsic $k_t$, initial (IPS) and initial+final state (IFPS) 
parton shower.~\cite{sampao}} 
\label{fig:fig2}
\end{center}
\end{figure}

  Fig.~\ref{fig:fig2}~\cite{sampao}    
  studies kinematic  corrections  to $B$-jet production   
due  to  longitudinal  momentum shifts in 
the initial-state parton shower, similar to those   discussed 
 for inclusive jets in  Fig.~\ref{fig:fig1}. 
For    $B$-jets in different rapidity regions~\cite{cms1202-b},  
 the  gluon $x$  distribution   is plotted 
 from   \powheg\  before  parton showering and after including various 
 components of  the parton  shower generator, using   the \pythia\  parton shower  
 Z2~\cite{tunes}    
 (including hadronization components to identify  $B$-jets). 
  Fig.~\ref{fig:fig2}  shows 
 similar shifts    in longitudinal momentum 
 with increasing  rapidity    as in the inclusive  jet case. 
A better understanding of $B$ production in this  region 
will affect 
 studies of the Higgs to $ b {\bar b}$ decay channel, e.g.  in the  
associated  production   with vector bosons. 

Although  the  explicit calculations in  Fig.~\ref{fig:fig2} are performed using 
  a  particular  NLO-shower  matching 
 scheme (\powheg),  the effect    is common  to any calculation matching NLO 
 with collinear showers. As discussed in~\cite{sampao,coki-1209}, the kinematic 
shifts due to the momentum reshuffling    can affect predictions of 
matched  NLO-shower    
calculations  both through  the perturbative weight for each event and through 
the  evaluation of the parton distribution functions. 
In    calculations  using integrated parton density functions 
 this implies that    correction factors as discussed in Sec.~2  
    have to be applied after the  evaluation of the cross section.  
On the other hand,  we note that  this is avoided  in approaches 
   using transverse momentum dependent  
    pdfs~\cite{jccbook,avsar11,mert-rog-11,unint09}  
     from the beginning   (TMDs or uPDFs).  
It will thus  be   of  interest to  
study   it   quantitatively in  Monte Carlo 
generators  which implement these  pdfs~\cite{cascade_docu,determ,jadach09}. 

We note further that helicity amplitudes  techniques 
are being developed~\cite{hameren1}   for multi-leg  processes  
with up to two  off-shell gluons 
in the high-energy limit~\cite{cch},   and applied to complex final 
states containing heavy quarks~\cite{hameren2}, including $ b {\bar b}  $ + jets and 
 $ b {\bar b}   $  +  vector bosons.  Once combined with TMD pdfs and 
 parton showers,  these  multi-parton amplitudes  
can  provide a computational  tool to  overcome the limitations of 
collinear approximations,  and   
take into account both kinematical  and dynamical 
effects associated with multiple-scale processes.

\section{Multi-parton interactions and energy flow variables}
\label{sec:under}

Besides   jet  cross sections, 
event shape variables are studied at the 
LHC~\cite{evshape-cms,evshape-atl} and used  to 
characterize   the  events'  energy flow  and   the 
structure of    multi-jet   final states.  Baseline descriptions 
 of  multi-jet events are  obtained 
  by methods  merging  parton showers and multi-leg 
hard matrix elements~\cite{hoeche11}. 
First  measurements of  LHC    
hadronic event shapes~\cite{evshape-cms}    
point to   parton showering  effects  dominating  over 
   contributions from   hard matrix elements   evaluated at high multiplicity.   
Jet shape variables  describing  
     the jet's  internal structure and the  energy flow  within a  jet 
   are  also   studied~\cite{jetshape-lhc}.  
   These observables are used in searches for 
   potential new physics signals from decays of  massive states in the  boosted 
   regime~\cite{alth-boost},  or away from it~\cite{gouz13}.  Besides 
jet substructure,  such  observables   are sensitive   to soft  physics effects,  including 
 underlying events, pile-up,  
  multiple parton interactions~\cite{bartal}.

\begin{figure}[htb]
\vspace{60mm}
\includegraphics{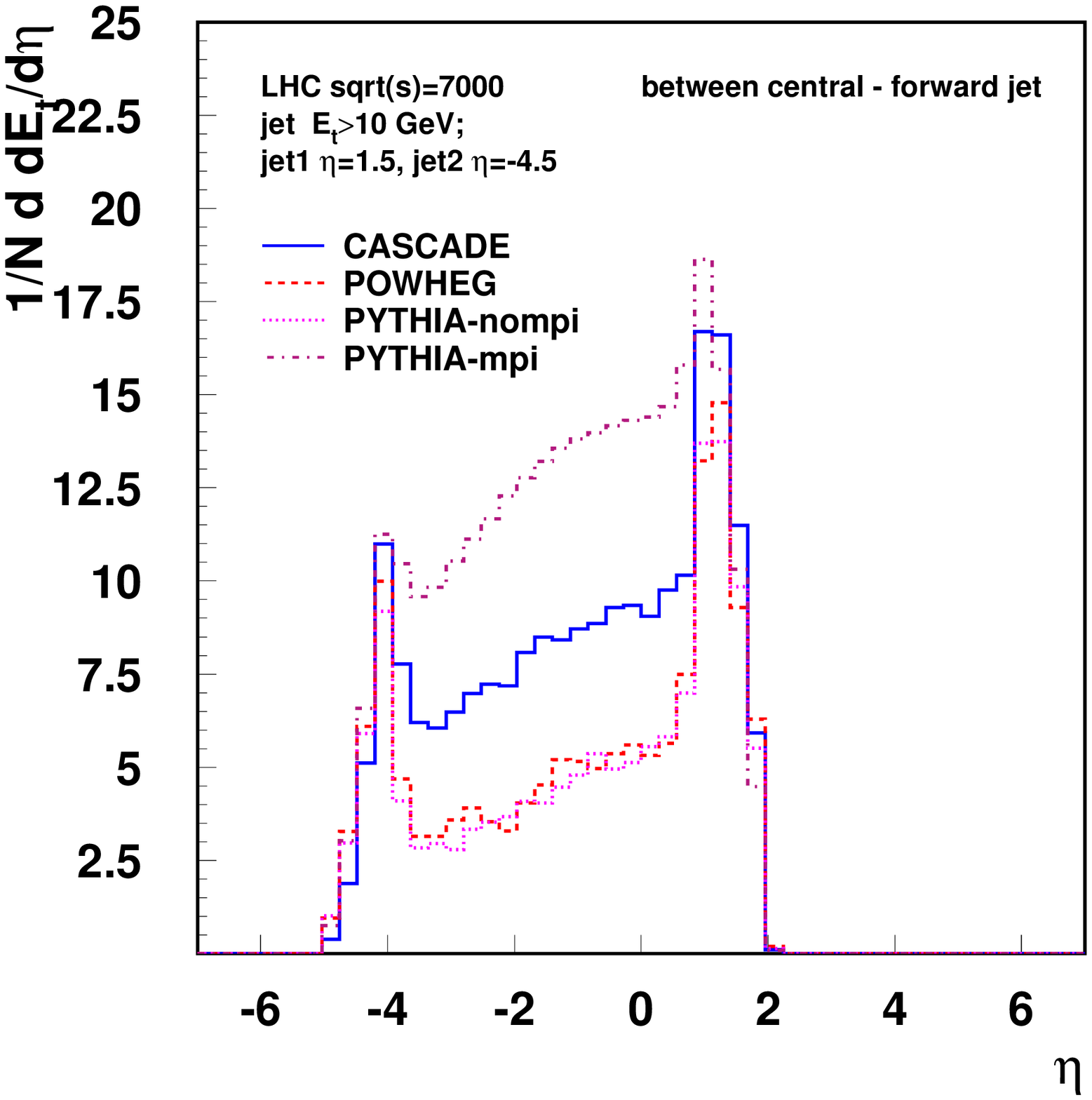}
\includegraphics{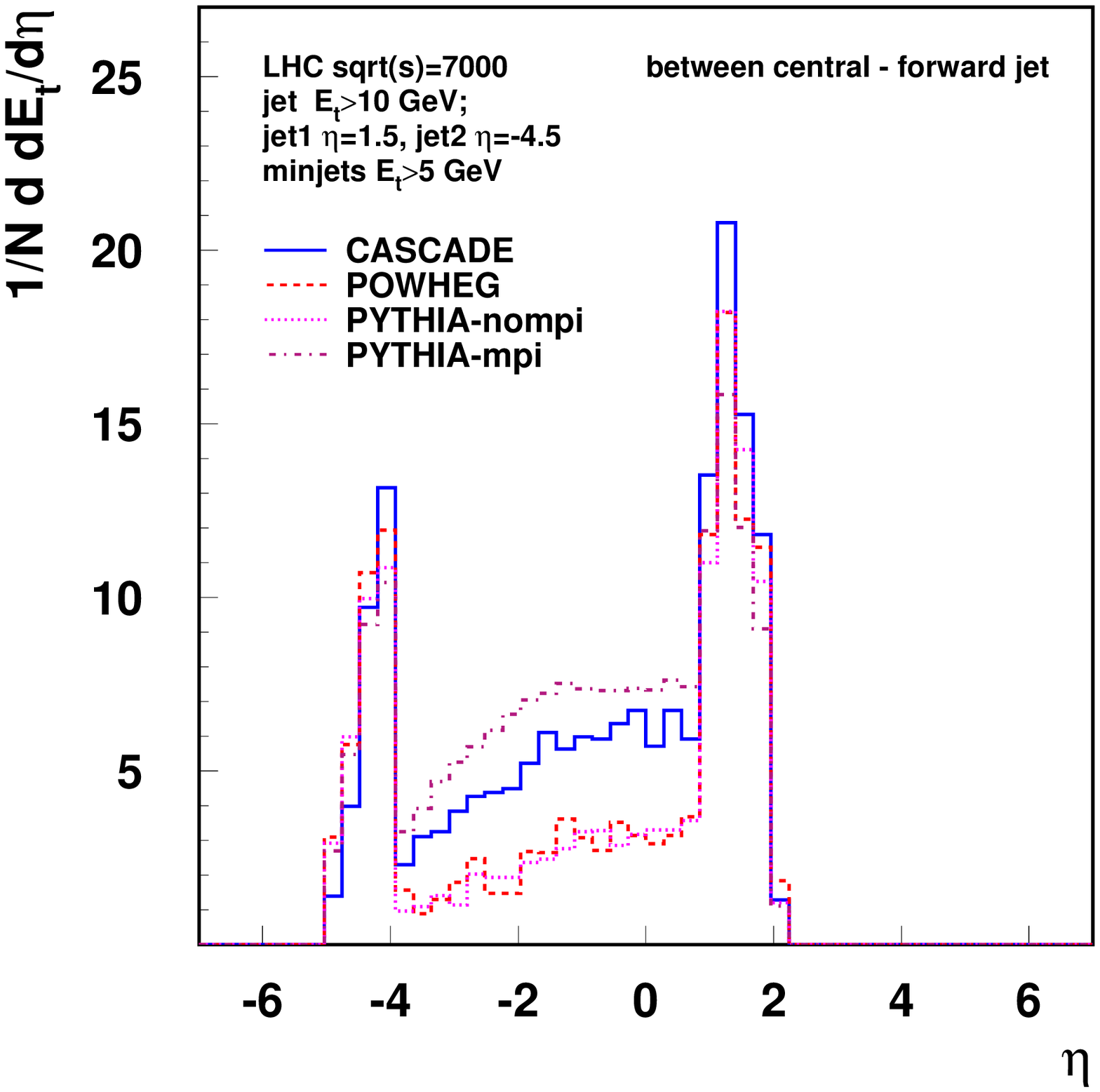}  
\caption{\it  Transverse 
energy flow~\cite{epjc12} in the  inter-jet   region:  (left) particle flow; (right) mini-jet flow. } 
\label{fig:betw} 
\end{figure}

 Energy flow measurements~\cite{cms-pas-10-02} 
  in minimum bias and dijet events,    
 designed to investigate 
properties of the soft underlying event, 
emphasize the  difficulty~\cite{bartfano1103}  in achieving   a unified 
underlying event description 
from central to forward rapidities, 
 based on \pythia~\cite{pz_perugia} Monte Carlo 
tuning.  Forward-backward correlations~\cite{skands11} 
 in minimum bias may help   analyze the 
event structure.   
Complementary to the above measurements 
are  transverse energy flow observables 
 associated with the production 
of  jets widely separated in rapidity~\cite{epjc12}, 
 sensitive to  harder color radiation, and useful for  
 studies   of showering and  of  multi-parton  interactions~\cite{eflow-may}.  
The transverse energy flow may be defined by 
summing  the energies over all particles 
in the final states above a  minimum $E_T$, or alternatively~\cite{epjc12} 
by first clustering   particles into jets  by means of a  jet algorithm, 
 and then  constructing    the associated energy flow from  jets  
with transverse energy 
 above a given lower bound $q_0$. In the latter case one  measures  
a (mini)jet energy flow, and 
 infrared safety is 
ensured by   the use of  the clustering  algorithm.

\begin{figure}[htb]
\vspace{53mm}
\includegraphics{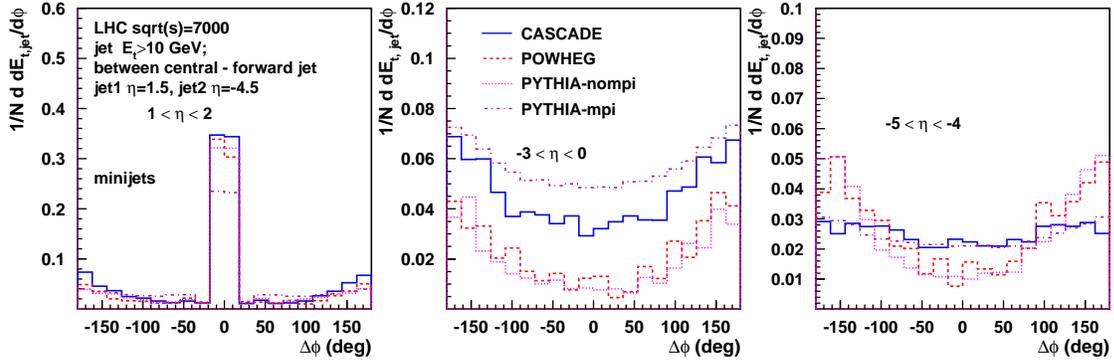}
\caption{\it  Azimuthal dependence of the  mini-jet   
energy flow~\cite{epjc12}   for different rapidity ranges:   
(left)  central-jet;   (middle) intermediate;  (right) forward-jet. 
  } 
\label{fig:azim} 
\end{figure}

Figs.~\ref{fig:betw}   and~\ref{fig:azim}
report results  for  the particle and minijet  energy   flow       
associated with production of central and forward jets~\cite{epjc12}  
 from three Monte Carlo event generators: 
 the  k$_\perp$-shower \cascade\  generator~\cite{cascade_docu},   to 
 evaluate contributions  of 
  high-energy logarithmic corrections; the NLO matched \powheg\  generator~\cite{alioli}, 
  to evaluate the  effects of  NLO   corrections to matrix elements; 
   \pythia\  Monte Carlo~\cite{pz_perugia},  
 used in two different modes:  with    the LHC tune  Z1~\cite{tunes}  
  (\pythia-mpi) to evaluate  contributions of 
   multi-parton interactions, 
and without  any  multi-parton interactions (\pythia-nompi).

Fig.~\ref{fig:betw}  shows the pseudorapidity dependence 
 of the  transverse energy flow  in the 
region between the central and forward jets. 
The particle energy flow   plot     on the left in Fig.~\ref{fig:betw}   
shows  the  jet profile picture,  and indicates     enhancements    of  
the energy flow   in the inter-jet region 
with respect to the \pythia-nompi    result 
 from higher order emissions in  \cascade\  and from multiple parton collisions in 
 \pythia-mpi. On the other hand, there is  little effect 
   from  the next-to-leading 
 hard correction  in \powheg\   with respect to  \pythia-nompi. 
  The energy flow  is dominated by
multiple-radiation, parton-shower effects.   The mini-jet energy flow  plot   on the right in 
Fig.~\ref{fig:betw}   
  indicates similar 
effects,   with reduced sensitivity  to infrared radiation. 
As the mini-jet flow definition suppresses the
contribution of soft radiation,    the  \cascade\  and  \pythia-mpi results become 
more similar in the inter-jet  region.  
Distinctive effects  are also found in~\cite{epjc12} by computations in the region away 
from the jets.

  Fig.~\ref{fig:azim}    illustrates  the azimuthal dependence 
of the mini-jet transverse energy flow. 
Here $\Delta \phi$   is  measured  with respect to the central jet.   The 
$\Delta \phi$  distribution is shown for three different rapidity ranges, 
corresponding  to the central-jet,  forward-jet,    and intermediate  rapidities. 
As we go  toward  forward rapidity,   the 
 \cascade\  and  \pythia-mpi calculations  give   a more   
 pronounced flattening  of the $\Delta \phi$  distribution compared 
 to  \powheg\   and   \pythia-nompi,    
  corresponding to   increased decorrelation between the jets. 

The above  numerical results  indicate that 
soft, finite-angle  multi-gluon emission   
over large rapidity intervals  gives  sizeable 
contribution to the inter-jet  energy flow. 
Non-collinear corrections  to single-chain 
showering may in particular  influence the rates for multi-parton interactions~\cite{bartal}.  
This also  underlines the relevance of approaches 
which aim at  a  more  complete   description of  
 initial state dynamics  by generalizing the 
notion of parton distributions~\cite{avsar11,mert-rog-11}, both 
for  quark-dominated~\cite{becher-neub} 
and gluon-dominated~\cite{xiao} processes. 
It will be  of interest to extend  energy flow studies  using   
different triggers, e.g.    
 vector boson + jet~\cite{martin-clara} at large rapidity, which   may help 
   constrain    multiple interaction rates~\cite{pz_perugia,tunes,ellie-paolo}.

\section{Towards  low $p_T$}
\label{sec:mini}

Hadron jets are measured at the LHC over 
 a wide range in transverse momenta from  20 GeV to a few TeV. 
It is of interest to explore to what extent 
the  picture of  jets  provided by the 
  algorithms used for jet reconstruction (and corresponding 
 shower Monte Carlo event simulations)  
 can be pushed towards  the semi-soft region of lower transverse 
momenta. As discussed earlier, 
 low-p$_T$  jets may  be used  to 
investigate QCD physics at small $x$.
A jet-like description of low-p$_T$ hadronic final states 
may  provide   insight into the    region 
where the LHC has produced striking evidence~\cite{ridge-pp-cms}  
of long-range  di-hadron correlations in high multiplicity events.

It has been observed  in~\cite{greb1209} that 
if    jet  measurements   at the LHC  are  extended 
down to   
 transverse momenta  of the order of a  few GeV 
  one can define a   visible  leading jet cross section   
  sensitive to 
the  unitarity 
 bound set  by the inelastic proton-proton rate  which  has recently been 
measured~\cite{Aad:2011eu,CMS-PAS-QCD-11-002,CMS-PAS-FWD-11-001}.  
This can be done  within the range of 
acceptance of the measurement without using any 
extrapolation~\cite{greb1209}.  Because of the low transverse momenta, 
this will   rely  primarily  on jets constructed from charged tracks. 
Given the decay in 
particle tracking capabilities   with increasing  rapidity, one may  
focus on the central  pseudorapidity  range. 

\begin{figure}[htbp]
\includegraphics[scale=0.35]{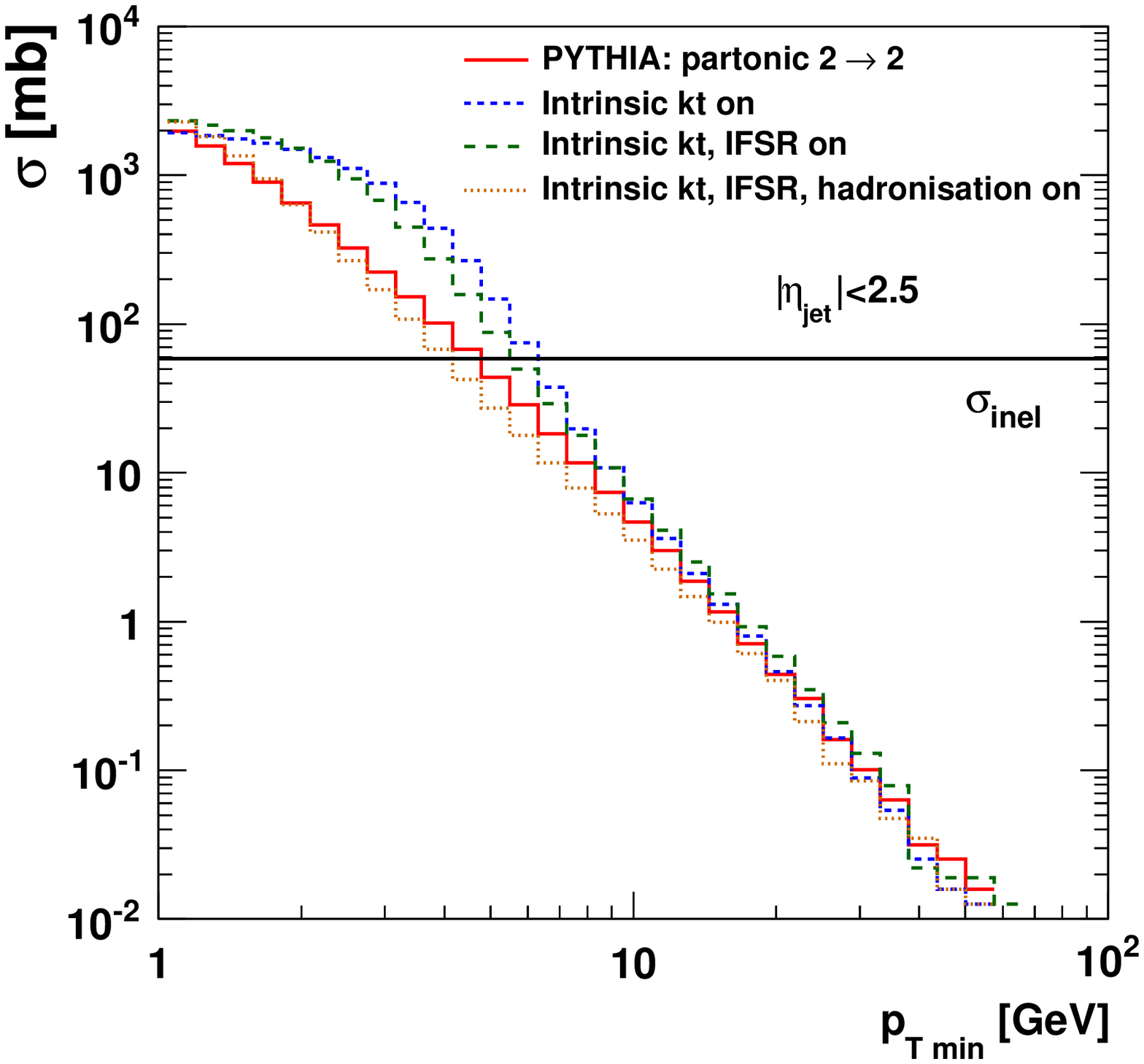}
\includegraphics[scale=0.35]{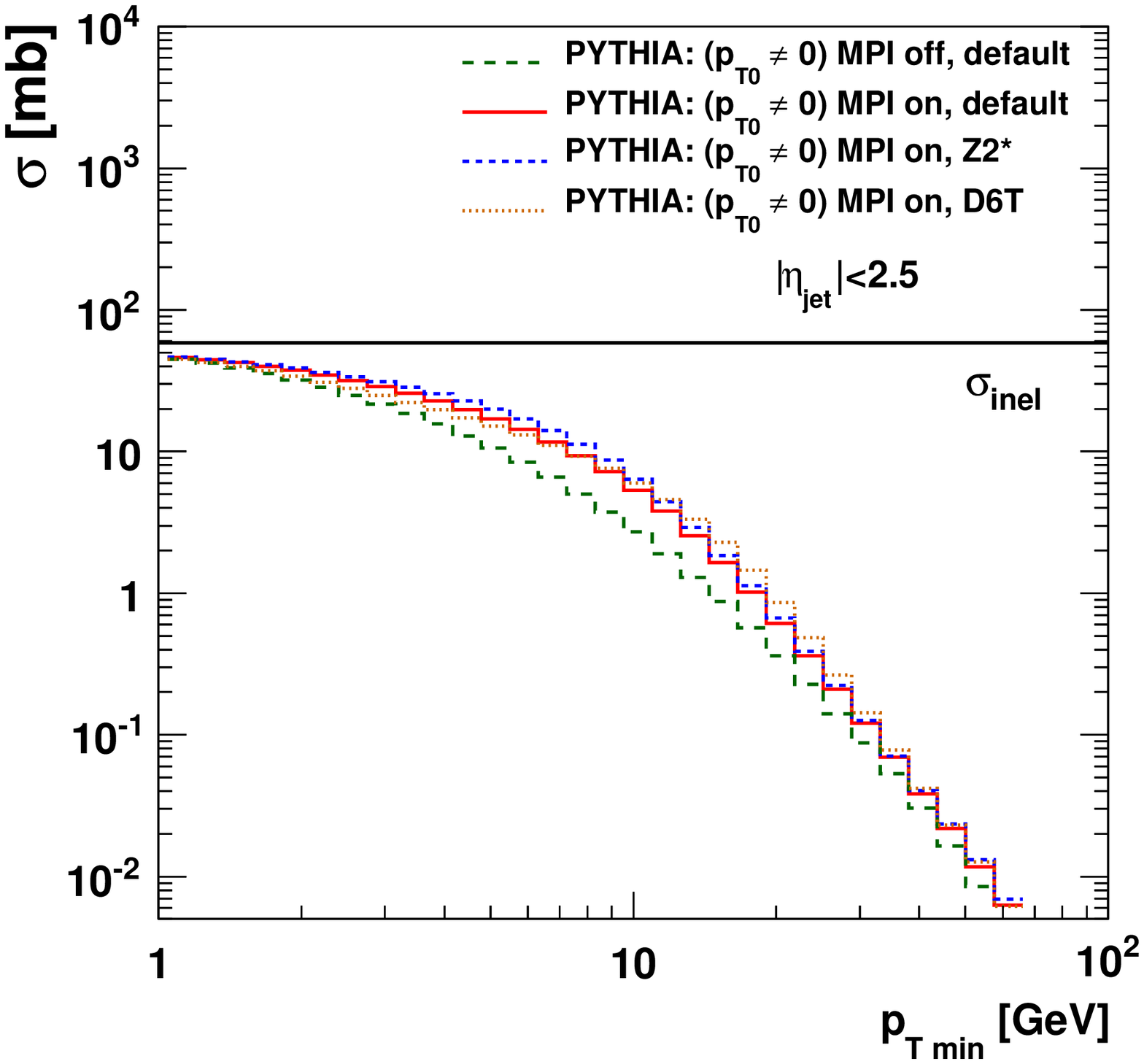}
\caption{\it (Left) Cross section from purely partonic $2\to2$ process, including intrinsic $k_t$-effects, including initial and final state parton showers (IFSR) and finally hadronisation.
(Right) predicted cross section applying $p_{T0}\neq 0 $ and MPI with different underlying event tunes of \protect\pythia .~\cite{greb1209} }  
\label{fig:fig2jetpp}
\end{figure}

The main interest of these measurements is the possibility to 
investigate    the leading jet cross section near the 
$p_T$  region, $p_T  = {\cal O } $ (a few GeV),   where 
 the inelastic pp production rate is saturated~\cite{greb1209}. 
 At the LHC, for the first time in collider experiments, 
this occurs in the weakly coupled region. 
Even though at weak  coupling, 
 dynamical effects slowing down the rise of the 
   cross section  in this region 
  involve strong fields and non-perturbative  physics. 
 In order to analyze   the saturation region, 
Ref.~\cite{greb1209} introduces an event cross section, defined as 
an integral over the differential leading jet cross section, 
which does not depend on the jet multiplicity.  
Fig.~\ref{fig:fig2jetpp}  shows the visible jet cross section 
using \pythia~\cite{pythref}  
 with  jets     reconstructed by    the  
anti-k$_T$    algorithm~\cite{antiktalgo}   
for  $R=0.5$ down to    low transverse momenta.

In Fig.~\ref{fig:fig2jetpp} (left)   the  perturbative result  reaches the 
inelastic bound~\cite{Aad:2011eu} for minimum $ p_T \simeq 4 $ GeV. 
In  the region  just above this value, $p_T  = {\cal O } (10)$ GeV,   effects  responsible for the taming of the cross section set in. 
  Fig.~\ref{fig:fig2jetpp} (right)  shows  the cross section based on 
the model~\cite{sjozijl,sjoska04},  in which  
 the rise of the cross section is tamed at small values of $p_T$ by introducing a  $p_{T0}$ cut-off 
parameter obtained from   fits  to describe measurements of the underlying event. 

\begin{figure}[htbp]
\begin{center} 
\includegraphics[scale=0.32]{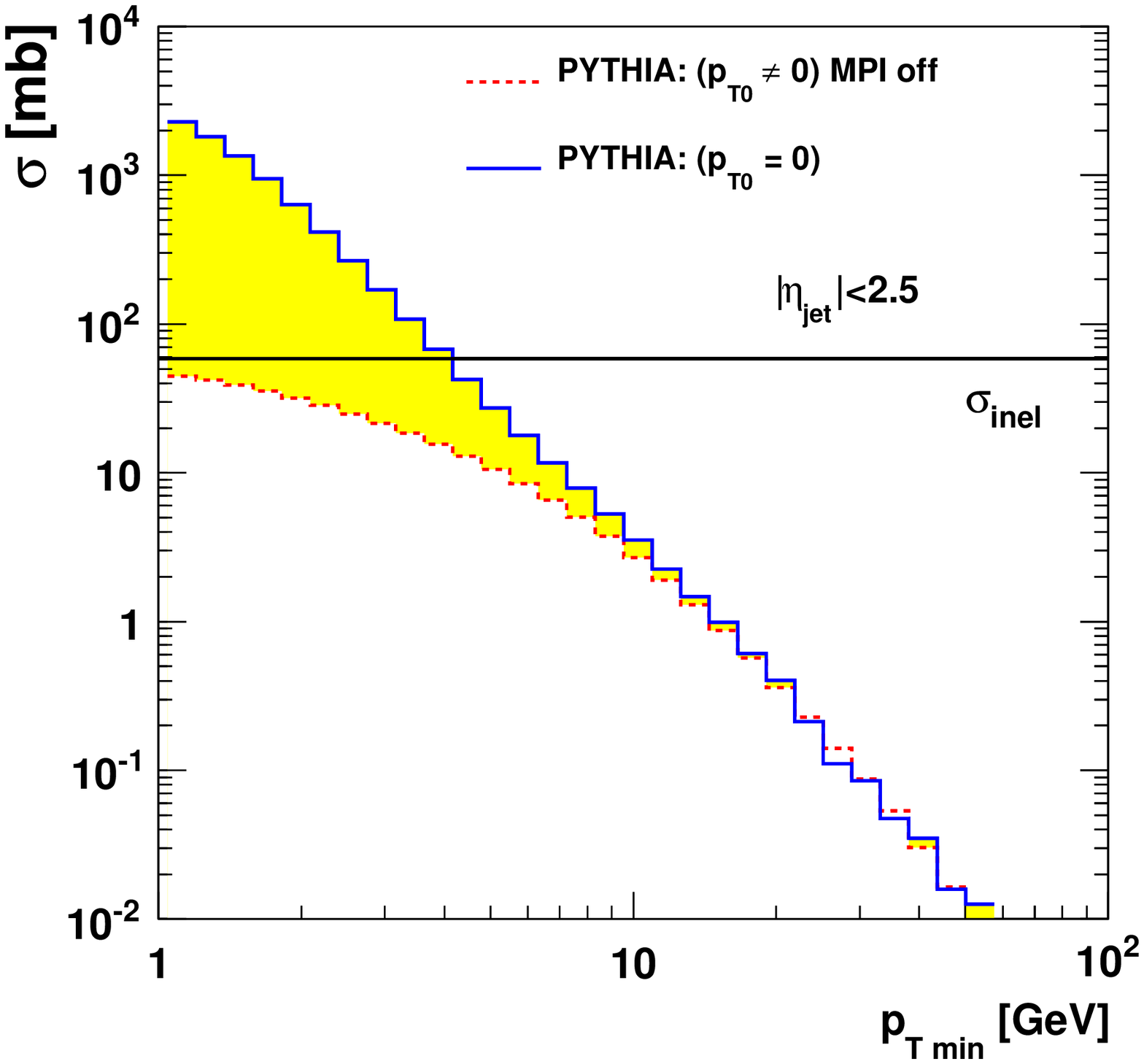}
\includegraphics[scale=0.32]{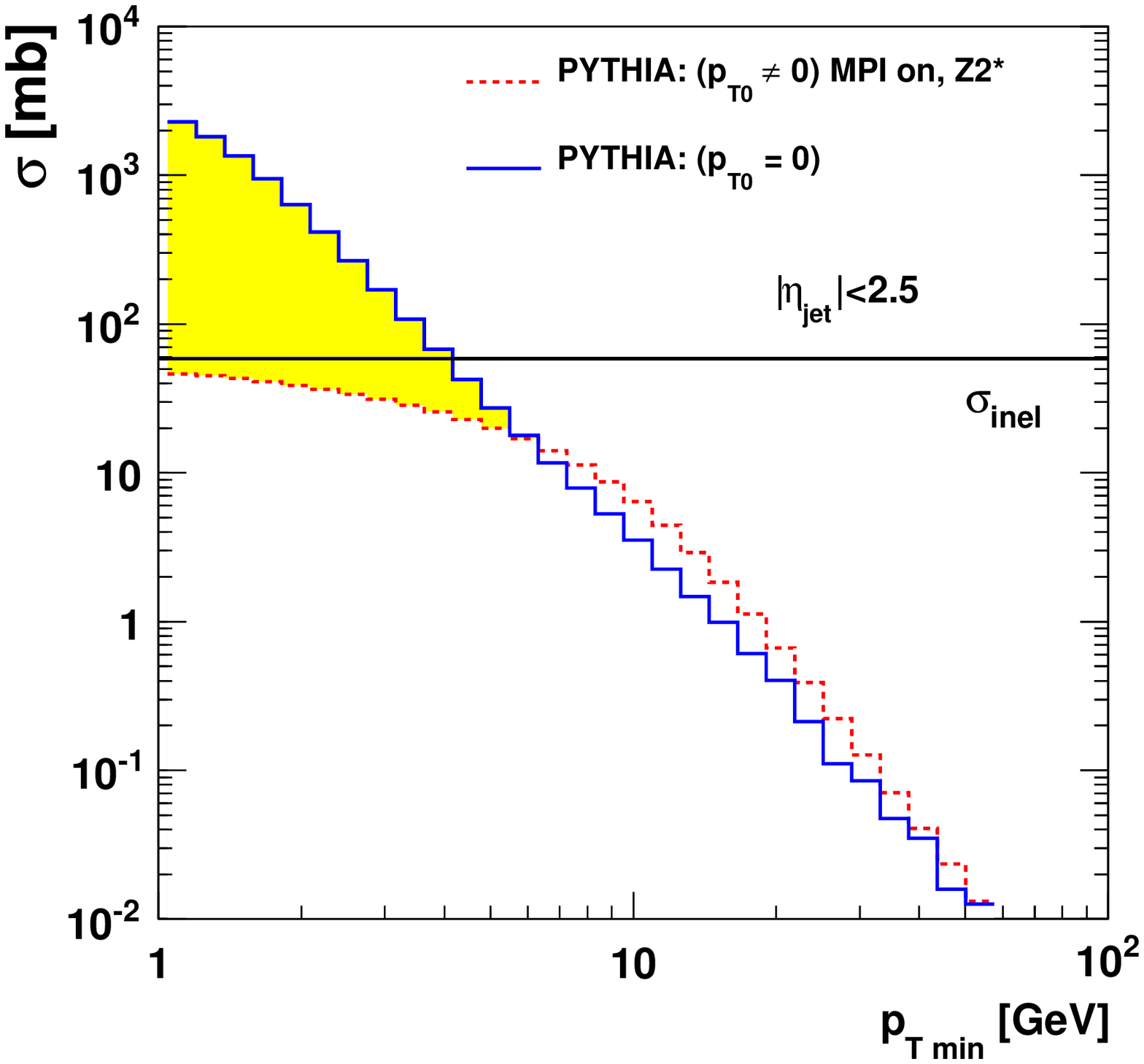}
\caption{\it The cross section~\cite{greb1209}   
as a function of $p_{T\,min}$ as predicted by  \protect\pythia ~in the range 
$|\eta| < 2.5$. The solid (blue) line shows the prediction applying
$p_{T0}=0$ including parton shower and hadronisation, while the dashed (red) line shows the prediction with $p_{T0}\neq 0$;
(left) is without multi-parton interactions, (right)  including multi-parton interactions with tune Z2* ~\protect\cite{tunes}.}
\label{fig:fig3mini}
\end{center} 
\end{figure} 

 Fig.~\ref{fig:fig3mini}~\cite{greb1209}   shows 
 a comparison of the jet cross section for $p_{T0}=0$, including parton shower and hadronization,  with the cross section obtained from \pythia\     including the $p_{T0}$  
model. In Fig.~\ref{fig:fig3mini} (right) we show the effect  of multi-parton interactions. Especially in the region of $p_T < 10$~GeV a clear deviation from the $p_{T0}=0$ prediction is visible. Measuring 
 the jet cross section in this range  would probe 
 the transition from the 
large-$p_T$ 
perturbative behavior   to the medium to low  $p_T$ region 
where (weak-coupling) unitarity corrections set in. We note that 
the  analysis~\cite{atlas-low-pt}  already   illustrates the 
feasibility of measuring jets at low $p_T$. However, 
 event cross sections such as that described above to study  unitarity 
have not yet been examined.  Results from  different 
  Monte Carlo models  in~\cite{atlas-low-pt}  are effectively 
normalized to the lowest $p_T$ bin. 

Besides $pp$ collisions,  low $p_T$  jet measurements are relevant 
 in pA and AA nuclear collisions. If the inelastic   
cross section is measured in AA and pA, they may be useful  
to characterize properties of final states in terms of jets or  
flows~\cite{werner-etal-12}, 
 and investigate  transverse momentum   
dependent dynamics    
and multi-parton  interactions~\cite{dent1211,mark-acta}.

 Compared 
 to the collinearly-factorized cut-off model~\cite{sjozijl,sjoska04}, 
 a physically  different  picture of the turn-over region 
is provided by approaches based on TMD 
initial-state distributions~\cite{avsar11,unint09,gosta}  
 incorporating small-$x$ gluon coherence~\cite{coh-x}. 
In this case the  rise of the cross section is slowed down by 
finite transverse momenta in the initial state.  The effect of 
combining this with nonlinear terms in the  evolution of 
 parton cascades~\cite{kutak-nonlin} 
and the dependence of the turn-over region on 
the infrared behavior of the strong  coupling~\cite{yuri04}
warrant   investigations.  Extending jet definition and 
 measurements to the region of  low transverse momenta  may  
 give us  radically new  information 
on the underlying physical picture   of high-energy multi-particle production.

\section{Conclusion}
\label{sec:conclu}

Jets  enter in essentially all areas of  the LHC physics  program,  
 from new particle  discovery  processes 
 to precision studies of the Standard Model to searches for physics 
 beyond the Standard Model.   Experimental results in the first three years 
of running of  the LHC 
  have  provided us  not 
 only  with solid confirmation  of our understanding of jet physics  from  QCD 
but also with challenges and  surprises.  
These involve both the high-p$_T$  region, featuring  several new effects of 
QCD parton showers, and  the low-p$_T$ region, probing   
nonperturbative components of  jet  cross sections.   
Much is to be learnt about the theory of hadronic jets and its  
  applications 
 to precision phenomenology as forthcoming LHC   data analyses  and 
measurements keep exploring  jets in  new  regions of phase space and probing 
different sectors of  QCD.

\vskip 0.6 cm 

\noindent  {\bf Acknowledgements}.\    The 
material presented in this article originated from 
discussion and collaboration with  
many   people, in particular  S.~Dooling,  D.~d'Enterria,  
 A.~Grebenyuk,  P.~Gunnellini,   M.~Deak,  H.~Jung,   M.~Hentschinski, 
N.~McCubbin,  Z.~Nagy,  
P.~Katsas, A.~Knutsson,   K.~Kutak,    K.~Rabbertz, C.~Roda.



\begin{footnotesize}

\end{footnotesize}

\end{document}